\documentclass[prd,twocolumn,showpacs,floatfix,superscriptaddress,nofootinbib]{revtex4}
\usepackage[utf8]{inputenc}
\usepackage{graphicx}
\usepackage{epsfig}
\usepackage{bm}
\usepackage{amsfonts}
\usepackage[T1]{fontenc}
\usepackage{amssymb}
\usepackage{float}
\usepackage{amsmath}
\usepackage{dcolumn}
\providecommand{\abs}[1]{\lvert#1\rvert}

\usepackage{cancel}
\usepackage[colorlinks]{hyperref}
\usepackage[usenames,dvipsnames]{color}
\hypersetup{
     breaklinks=true,
    pdfstartview={FitH},    % fits the width of the page to the window
    colorlinks=true,       % false: boxed links; true: colored links
    linkcolor=blue,          % color of internal links
    citecolor=red,        % color of links to bibliography
    filecolor=magenta,      % color of file links
    urlcolor=blue,           % color of external links
    anchorcolor=green,      % Color for anchor text
    linktocpage=true
}

\def\doi{http://doi.org}

\def\be{\begin{equation*}}
\def\ee{\end{equation*}}

\begin{document}

\title{Study of Null Geodesics and their Stability in Kalb-Ramond Black  Holes }

\author{P. A. Gonz\'{a}lez}
\email{pablo.gonzalez@udp.cl} \affiliation{Facultad de
Ingenier\'{i}a y Ciencias, Universidad Diego Portales, Avenida Ej\'{e}rcito
Libertador 441, Casilla 298-V, Santiago, Chile.}
\author{Marco Olivares}
\email{marco.olivaresr@mail.udp.cl}
\affiliation{Facultad de
Ingenier\'{i}a y Ciencias, Universidad Diego Portales, Avenida Ej\'{e}rcito
Libertador 441, Casilla 298-V, Santiago, Chile.}
\author{Eleftherios Papantonopoulos}
\email{lpapa@central.ntua.gr}
\affiliation{Physics Division, School of Applied Mathematical and Physical Sciences, National Technical University of Athens, 15780 Zografou Campus,
    Athens, Greece.}
\author{Yerko V\'{a}squez}
\email{yvasquez@userena.cl}
\affiliation{Departamento de F\'{\i}sica, Facultad de Ciencias, Universidad de La Serena,\\
Avenida Cisternas 1200, La Serena, Chile.}
\date{\today}

\begin{abstract}
We investigate null geodesics and their stability in four-dimensional charged Kalb--Ramond black holes with a cosmological constant. The non-vanishing vacuum expectation value of the background antisymmetric tensor field induces spontaneous Lorentz symmetry breaking, controlled by a dimensionless parameter $l$, and deforms the Reissner--Nordstr\"om--(A)dS geometry. We derive the effective potential for massless particles and obtain analytic expressions for the photon-sphere radius, critical impact parameter, capture cross section, and deflection trajectories. The limiting cases $l\rightarrow0$ recover the RN, RN--dS and RN--AdS geodesic structures, while finite values of $l$ shift the turning points, modify the bending of light and deform both first-kind and second-kind photon trajectories. In the AdS branch, we find a special limiting lima\c{c}on-type null geodesic whose angular structure is modified by the Kalb--Ramond parameter. We also compute the Lyapunov exponent of the unstable circular null orbit and show that the cosmological constant changes the instability timescale, whereas the Lorentz--violating parameter produces a non-trivial deformation of the photon-sphere instability. Our results identify null geodesics as sensitive probes of Lorentz--violating effects in charged black hole spacetimes.
\end{abstract}

\maketitle

%\printindex

\tableofcontents

\section{Introduction}

General Relativity (GR) provides an accurate description of gravitational phenomena across a wide range of scales, from Solar System experiments to the dynamics of compact objects and gravitational waves. Nevertheless, it is widely expected that GR should arise as an effective description of a more fundamental theory, and that deviations from it may become relevant in high energy or strong field regimes. In this context, string-inspired effective theories generically contain, in addition to the metric tensor, higher rank tensor fields. A prominent example is the Kalb--Ramond (KR) field $B_{\mu\nu}$, an antisymmetric rank-two tensor whose field strength is a three-form and which naturally appears in the low energy effective action of string theory \cite{Kalb:1974yc}.

When the KR field is non-minimally coupled to gravity, it can induce non trivial modifications of the spacetime geometry. In particular, if the KR field acquires a non vanishing vacuum expectation value, Lorentz symmetry may be spontaneously broken. Lorentz symmetry breaking refers to the violation of invariance under local rotations and boosts, implying the existence of preferred directions in spacetime. This mechanism has been extensively discussed in string-inspired scenarios and in effective descriptions of Lorentz--violating gravity \cite{Kostelecky:1989jw,Bluhm:2004ep}. At the cosmological level, spontaneous symmetry breaking at the Planck scale can leave relic tensor backgrounds permeating the Universe, potentially generating anisotropies or modifying the evolution of cosmological perturbations. In particular, Lorentz--violating background fields have been shown to affect late time cosmological dynamics and may give rise to de Sitter attractor solutions, as well as to instabilities capable of amplifying perturbations on scales comparable with the present Hubble radius \cite{Libanov:2007mq}.

Black hole solutions in the presence of a background KR field have attracted increasing attention in recent years. Static and spherically symmetric solutions were obtained in gravitational theory, where the Lorentz symmetry is spontaneously broken by the vacuum expectation value of the KR field \cite{Yang:2023mbr}. These geometries are characterized by a dimensionless Lorentz--violating parameter, usually denoted by $l$, which modifies the metric functions and affects the horizon structure, thermodynamics, and classical gravitational observables. Subsequently, charged extensions were constructed in both the absence and in the presence of a cosmological constant \cite{Duan:2023qsg}. The resulting charged KR black holes take a Reissner--Nordstr\"om-like form deformed by the Lorentz--violating background, thereby providing a natural arena to test how charge, cosmological constant, and Lorentz symmetry breaking jointly affect black hole physics.

Several aspects of KR black holes have already been investigated. The thermodynamic properties of the neutral and charged configurations were analyzed in Refs.~\cite{Yang:2023mbr,Duan:2023qsg}, showing that the KR parameter modifies the local stability range and the phase structure. The optical properties of these geometries have also been studied through gravitational lensing, shadows, and ray-tracing techniques. In particular, exact and approximate deflection angles have been obtained for Schwarzschild-like KR black holes, while charged KR black holes have been analyzed in terms of their photon sphere, critical impact parameter, and optical appearance in the presence of thin accretion disks \cite{Junior:2024vdk,Tan:2025pya}. Further developments include weak deflection angles, shadows, quasinormal modes, neutrino annihilation, plasma effects, and different accretion models \cite{Mangut:2025gie,Pantig:2025eda,Xu:2025iwg}. These studies indicate that the Lorentz--violating KR parameter can leave imprints on potentially observable strong field phenomena.

The perturbative sector and physically motivated extensions of KR black holes have also been explored. Quasinormal modes and greybody factors of Lorentz--violating black holes were studied in~\cite{Guo:2023nkd}, while electrically charged and slowly rotating KR black holes have been analyzed through their quasinormal spectra, shadows, and perturbative dynamics \cite{Gu:2025lyz,Liu:2024lve,Deng:2025atg}. Black holes influenced by a global monopole charge in KR gravity were investigated in Ref.~\cite{Baruah:2025ifh}, where quasinormal modes, greybody factors, and the sparsity of Hawking radiation were studied. Further developments include KR black holes surrounded by quintessence fields, perfect-fluid dark matter, nonlinear electrodynamics and dyonic charges, where geodesic dynamics, thermodynamic criticality, optical appearance, ringdown signatures, shadows, and Hawking radiation sparsity have been investigated \cite{Al-Badawi:2025amp,Liang:2026gjv,Ahmed:2026nzy,Lin:2026ewo}. In addition, Lorentz--violating signatures in quasi-periodic oscillations from magnetized KR black holes have recently been proposed as a possible observational probe \cite{Rodrigues:2026ofx}. These results suggest that KR gravity provides a useful framework for investigating departures from GR in the classical, semiclassical, and observational aspects of black hole physics.

A fundamental question in this context concerns how the loss of Lorentz invariance affects the motion of the test particles. Geodesics provide a direct probe of the spacetime structure and allow one to identify possible observational signatures of the underlying theory. In particular, null geodesics are central to black hole phenomenology, since they determine the photon sphere, critical impact parameter, capture cross section, bending of light, and optical appearance of compact objects. They are also related to time delay effects, gravitational redshift, and the instability timescale of circular photon orbits. Therefore, the study of null geodesics provides a direct way to quantify the impact of the KR Lorentz--violating parameter on light propagation in strong gravitational fields.

The inclusion of a cosmological constant makes the geodesic structure richer. In asymptotically de Sitter spacetimes, the cosmological horizon restricts the allowed region of photon propagation and modifies the global causal structure. In asymptotically anti-de Sitter spacetimes, the confining character of the geometry allows for second kind trajectories and special analytic orbits that are absent in asymptotically flat backgrounds. Consequently, charged KR black holes with $\Lambda=0$, $\Lambda>0$ and $\Lambda<0$ provide a natural setting to compare how Lorentz symmetry breaking affects photon motion under different asymptotic conditions. In this setting, the limit $l\to0$ plays a central role. 
It provides the Reissner--Nordstr\"om (RN), Reissner--Nordstr\"om--de Sitter (RN--dS) and Reissner--Nordstr\"om--anti-de Sitter (RN--AdS) geometries as reference backgrounds. 
Therefore, the charged KR solution can be understood as a controlled Lorentz-violating deformation of the RN--(A)dS family. 
This perspective allows one to identify which features of photon motion are inherited from the standard charged black-hole geometry and which ones are genuine consequences of the KR background.

Motivated by these considerations, in this work we analyze the null geodesic structure of four-dimensional charged KR black hole spacetimes in the presence of a cosmological constant. 
We derive the effective potential for massless particles and obtain analytic expressions for the main quantities governing photon motion, including the photon-sphere radius, the critical impact parameter, the capture cross section and the deflection trajectories. 
Throughout the analysis we use the $l=0$ limit as a benchmark, so that the effects of the KR parameter can be interpreted as deviations from the corresponding RN, RN--dS and RN--AdS cases. 
We study radial and angular trajectories, including light deflection, critical trajectories, second-kind trajectories, capture regions and the transition between deflection and plunging regimes. 
We also discuss weak-field observables, such as gravitational redshift and Shapiro time delay, and characterize the instability of unstable circular null orbits through the corresponding Lyapunov exponent.

We show that the Lorentz-violating KR parameter modifies the effective gravitational potential and changes the characteristic scales associated with photon propagation. 
In particular, it shifts the unstable circular photon orbit, changes the critical impact parameter and affects the capture and deflection sectors. 
By contrast, the cosmological constant does not shift the photon-sphere radius, but changes the energetic thresholds and the global structure of the trajectories. 
Thus, the KR parameter mainly controls the local deformation of the photon-sphere and effective-barrier structure, whereas the cosmological constant distinguishes the asymptotically flat, de Sitter and anti-de Sitter sectors. 
This separation of effects provides a useful way to identify Lorentz-violating signatures in black-hole optical phenomena.

The paper is organized as follows. In Sec.~\ref{FDKR} we review the four-dimensional charged KR black hole geometry and discuss the role of the Lorentz--violating parameter and the cosmological constant. In Sec.~\ref{NGS} we perform a detailed analysis of null geodesics, including radial
motion, angular trajectories, bending of light, critical orbits,
and associated capture and deflection regions.  In Sec.~\ref{OBE} we discuss weak-field observational effects associated with photon propagation, namely bending of light, gravitational redshift and Shapiro time delay. 
In Sec.~\ref{LE} we compute the Lyapunov exponent associated with unstable null circular orbits. Finally, in Sec.~\ref{conclution} we summarize our results and discuss their physical implications.

\section{Four-Dimensional Kalb--Ramond Black Holes}
\label{FDKR}

We consider a four-dimensional gravitational theory in the presence of a
KR field, described by a rank-two antisymmetric tensor
$B_{\mu\nu}=-B_{\nu\mu}$. The dynamics is governed by the action \cite{Duan:2023gng}
\begin{eqnarray}
\notag S&=&\frac{1}{2}\int d^4x\,\sqrt{-g}\Big[
R-2\Lambda
-\frac{1}{6}H_{\mu\nu\rho}H^{\mu\nu\rho}\\
\notag && -V(B_{\mu\nu}B^{\mu\nu}\pm \bar{b}^2)
+\xi_2 B_{\rho\mu}B^{\nu\mu}R^{\rho}{}_{\nu}\\
&& +\xi_3 B_{\mu\nu}B^{\mu\nu}R
\Big]
+\int d^4x\,\sqrt{-g}\mathcal{L}_M\,,
\label{actionKR}
\end{eqnarray}
where
\begin{equation}
H_{\mu\nu\rho}=\partial_{[\mu}B_{\nu\rho]}\,,
\end{equation}
is the KR field strength, $\Lambda$ is the cosmological constant, and
$\xi_2,\xi_3$ denote the non-minimal coupling constants between the KR
field and gravity.

To support electrically charged black hole solutions, the matter sector is
taken as
\begin{equation}
\mathcal{L}_{\text{M}}
=
-\frac{1}{2}F_{\mu\nu}F^{\mu\nu}
-\eta B_{\alpha\beta}B_{\gamma\rho}F^{\alpha\beta}F^{\gamma\rho} \,,
\label{matterKR}
\end{equation}
where $F_{\mu\nu}=\partial_\mu A_\nu-\partial_\nu A_\mu$ is the Maxwell
tensor and $\eta$ is the coupling constant between the electromagnetic and
KR sectors.

The potential $V(B_{\mu\nu}B^{\mu\nu}\pm \bar{b}^2)$ induces spontaneous Lorentz
symmetry breaking through a non-vanishing vacuum expectation value of the
KR field,
\begin{equation}
\langle B_{\mu\nu}\rangle=b_{\mu\nu}\,,
\end{equation}
with
\begin{equation}
b_{\mu\nu}b^{\mu\nu}=\mp \bar{b}^2\,,
\end{equation}
where $\bar{b}^2$ is a positive constant. In the static and spherically symmetric
sector, one may consider a background configuration such that the KR field
strength vanishes, $H_{\mu\nu\rho}=0$, while the non-minimal couplings still
leave a non-trivial imprint on the geometry.

Under these assumptions, the theory admits static and spherically symmetric
black hole solutions of the form \cite{Duan:2023gng}
\begin{equation}
ds^2=-f(r)\,dt^2+\frac{dr^2}{f(r)}+r^2d\Omega^2\,,
\label{metricKR}
\end{equation}
with metric function
\begin{equation}
f(r)=\frac{1}{1-l}-\frac{2M}{r}
+\frac{Q^2}{(1-l)^2r^2}
-\frac{\Lambda r^2}{3(1-l)}\,,
\label{FrKRLambda}
\end{equation}
where
\begin{equation}
l=\frac{\xi_2 \bar{b}^2}{2}\,
\end{equation}
is the dimensionless Lorentz--violating parameter associated with the KR background.

%\end{equation}

This geometry reduces to the RN--(A)dS spacetime in the
limit $l\to0$. In turn, for $Q=0$ it reproduces a Schwarzschild--(A)dS-like
solution with Lorentz-violating corrections, whereas for $\Lambda=0$ one
recovers the asymptotically non-Minkowskian charged KR black hole.

It is convenient to note that the Lorentz--violating parameter modifies the
effective asymptotic structure of the spacetime. In particular, for
$\Lambda\neq0$, the large-$r$ behavior is controlled by
\begin{equation}
f(r)\sim -\frac{\Lambda}{3(1-l)}\,r^2\,,
\end{equation}
which shows that the effective cosmological scale is rescaled by the KR
background. Likewise, the electric contribution is modified through the
combination $Q^2/(1-l)^2$. From the lapse function and depending on the value of the cosmological constant $\Lambda$, we can study the location of the horizons by analyzing the three different configurations separately:

\begin{enumerate}
%\textcolor{red}{
\item Asymptotically flat Kalb-Ramond black hole $(\Lambda=0)$: 

The spacetime allows two horizons (the event horizon $\rho_{+}$ and the Cauchy horizon $\rho_{-}$)
\begin{equation}
\rho_{+}=M(1-l)+\sqrt { M^2(1-l)^2-{Q ^2\over (1-l)} }\,,
\label{g2.1}
\end{equation}
and
\begin{equation}
\rho_{-}=M(1-l)-\sqrt {M^2(1-l)^2-{Q ^2\over (1-l)} }\,,
\label{g2.1b}
\end{equation}

	\item   Kalb-Ramond  --de Sitter black hole $(\Lambda>0)$:
    
 The spacetime allows three horizons, the event horizon $R_+$, the cosmological horizon $R_{++}$, and the Cauchy horizon $R_-$, which are obtained from the quartic equation
    \begin{equation} 
        r^{4}-\frac{3}{\Lambda}r^2+\frac{6M(1-l)}{\Lambda}r-\frac{3Q^2}{(1-l)\,\Lambda }=0\,.
        \label{1.5}
     \end{equation} 
Its solutions are %\cite{COV}
    \begin{eqnarray}
          R_{++}&=&\alpha_{\Lambda} +\sqrt{-{3M (1-l)\over 2\Lambda\,\alpha_{\Lambda}}+{3 \over 2\Lambda}-\alpha_{\Lambda}^2}\,,\\
         R_{+}&=&\alpha_{\Lambda} -\sqrt{-{3M (1-l)\over 2\Lambda\,\alpha_{\Lambda}}+{3 \over 2\Lambda}-\alpha_{\Lambda}^2}\,,\\
        R_{-}&=&-\alpha_{\Lambda} +\sqrt{{3M(1-l) \over 2\Lambda\,\alpha_{\Lambda}}+{3 \over 2\Lambda}-\alpha_{\Lambda}^2}\,,\\
        R_{4}&=&-\alpha_{\Lambda} -\sqrt{{3M(1-l) \over 2\Lambda\,\alpha_{\Lambda}}+{3 \over 2\Lambda}-\alpha_{\Lambda}^2}\,,
    \end{eqnarray}
    where  
    \begin{eqnarray}
        \alpha_{\Lambda} &=&\sqrt{U_{\Lambda}\cosh \left[ \frac{1}{3}\cosh^{-1} \Xi_{\Lambda}\right]+\frac{1}{2\Lambda}}\,,\\
        U_{\Lambda}&=&{1 \over 2\Lambda}\sqrt{1-{4Q^2 \Lambda \over 1-l}} \,,\\
    \Xi_{\Lambda}&=&\frac{\left(18M^2\Lambda(1-l)^3-12Q^2\Lambda-1+l\right)\sqrt{1-l}}{(1-l-4Q^2 \Lambda)^{3/2}}\,.
    \end{eqnarray}
    The root $R_{4}$ is a negative solution  with no physical interpretation.
 
\item   Kalb-Ramond  --anti-de Sitter black hole $(\Lambda=-\frac{3}{\ell^{2}}<0)$: 

The spacetime allows two horizons (the event horizon $r_+$, and the Cauchy horizon $r_{-}$), which must be the real positive solutions to the quartic equation
	\begin{equation} r^{4}+\ell^{2}r^2-2M\ell^{2}(1-l) \,r+{Q^{2}\ell^{2}\over 1-l}=0 
  \,.  \label{1.3}\end{equation}
	 Its solutions are 
     %\cite{Cruz:2004ts}
	\begin{eqnarray}
	 r_{+}&=&\alpha_{\ell}
            +\sqrt{{M\ell^2 (1-l)\over 2\,\alpha_{\ell}}
            -{\ell^2 \over 2}
            -\alpha_{\ell}^2}\,,\\
        	r_{-}&=&\alpha_{\ell} -\sqrt{{M\ell^2 (1-l)\over 2\,\alpha_{\ell}}-{\ell^2 \over 2}-\alpha_{\ell}^2}\,,\\
	        r_{3}&=&-\alpha_{\ell} +\sqrt{-{M\ell^2 (1-l)\over 2\,\alpha_{\ell}}-{\ell^2 \over 2}-\alpha_{\ell}^2}\,,\\
	        r_{4}&=&-\alpha_{\ell} -\sqrt{-{M\ell^2 (1-l)\over 2\,\alpha_{\ell}}-{\ell^2 \over 2}-\alpha_{\ell}^2}\,,
	\end{eqnarray}
	where  
	\begin{eqnarray}
	\alpha_{\ell} &=&\sqrt{U_{\ell}\cosh \left[ \frac{1}{3}\cosh^{-1} \Xi_{\ell}\right]-\frac{\ell^2}{6}}\,,\\
	U_{\ell}&=&{\ell^2 \over 6}\sqrt{1+{12Q^2  \over (1-l)\,\ell^2}}\,, \\
	\Xi_{\ell}&=&\frac{(54M^2(1-l)^3+(1-l)\ell^2-36Q^2)\sqrt{(1-l)\,\ell^2}}{\left((1-l)\,\ell^2+12Q^2\right)^{3/2}}\,.
	\end{eqnarray}
        Through the discriminant for quartic polynomials, we can find in general that this is negative for all non-zero real $\ell$, so these four roots are distinct; on the one hand $r_+$ and $r_-$ are positive 
        real roots, respectively, while $r_3$ and $r_4$ are the complex conjugates of each other.    
\end{enumerate}

\section{Null Geodesic structure}
\label{NGS}

In order to study the motion of test particles in the background of the black  hole (\ref{metricKR})
we use the standard Lagrangian approach \cite{Chandrasekhar:579245,Cruz:2004ts,Villanueva:2018kem}.
The corresponding Lagrangian is

\begin{equation}
2\mathcal{L} =-f(r)\, \dot{t}^{2}+%
\frac{\dot{r}^{2}}{ f(r) }%
+r^{2}\left( \dot{\theta}^{2}+\sin ^{2}\theta \,\dot{\phi}^{2}\right) =-m^2\,.
\label{g6}
\end{equation}%
Here, the dot refers to the derivative with respect to an affine parameter $\lambda $,
along the trajectory, and, by normalization, $m=0\,(1)$ for massless (massive)
particles.
Since $(t, \phi)$ are cyclic coordinates, their corresponding
conjugate momenta $(\Pi _{t}, \Pi _{\phi })$ are conserved
and given by

\begin{equation}\label{g7}
\Pi _{t}=-f(r)\, \dot{t}=-E\,,
\qquad
\Pi _{\phi }=r^{2}\sin ^{2}\theta \,\dot{\phi}=L\,.
\end{equation}%
Therefore, considering that the motion is performed in an invariant plane, which
we fix at $\theta =\pi/2$, we obtain the
following expressions

\begin{equation}\label{g8}
\dot{t}=\frac{E}{f(r) }\,,%
\qquad\dot{\phi}=\frac{L}{r^{2}}\,.
\end{equation}

These relations together with Eq.(\ref{g6}) allow us to obtain the following
differential equations

\begin{eqnarray}\label{g9}
    \left(\frac{dr}{d\lambda}\right)^{2}&=&E^2 - V_{\text{eff}}(r)\,,\\ \label{g10}
    \left(\frac{dr}{dt}\right)^{2}&=&\frac{f(r)^2}{E^2}
    \,\left(E^2 - V_{\text{eff}}(r) \right)\,, \\ \label{g11}
    \left(\frac{dr}{d\phi}\right)^{2}&=&\frac{r^{4}}{L^2}\left( E^2-V_{\text{eff}}(r) \right)\,,
\end{eqnarray}%
where the effective potential $V_{\text{eff}}\left( r\right) $, reads

\begin{eqnarray}
\notag V_{\text{eff}}\left( r\right) &=&\left(\frac{1}{1-l}-\frac{2M}{r}
+\frac{Q^2}{(1-l)^2r^2}
-\frac{\Lambda r^2}{3(1-l)}\right)\times \\
&&\left( m^2+\frac{L^{2}}{r^{2}}\right)\,.  \label{g12}
\end{eqnarray}%
In the next sections, based on this effective potential, we analyze the
null geodesic structure ($m=0$) of the space-time characterized by the metric (\ref{FrKRLambda}).

\subsection{Motion with $L=0$}

The motion of photons with vanishing angular momentum $(L=0)$ is described by a null effective potential, $V_{\rm eff}=0$. Therefore, photons in this case can either escape to spatial infinity or plunge into the horizon. Thus, radial Eq. (\ref{g9}) reduces
to

\begin{equation}\label{n3}
\frac{dr}{d\lambda}=\pm E\,,
\end{equation}%
so, an elemental integration yields
\begin{equation}\label{n4}
\lambda\left( r,E\right) =\pm \frac{r-\rho_{0}}{E}\,.
\end{equation}%
Here, $\rho_{0}$ denotes the initial radial position of the massless particle. Note that the motion depends on the energy of the massless particle, and does not depend on the metric  function; thereby the above equation is valid for the three cases considered, that is, asymptotically flat, dS and AdS KR spacetimes. With respect to the
affine parameter the photons arrive at the horizon in a finite
affine parameter, and when the photons move in the opposite
direction, they require an infinite affine parameter to reach infinity. Note that when the energy of the photon increases the affine parameter decreases. 

\begin{enumerate}
%\textcolor{red}{
	\item Asymptotically flat  KR black hole  $(\Lambda=0)$:

By integrating Eq. (\ref{g10}) the coordinate time  yields 
\begin{eqnarray}\label{n5}
\notag t\left( r\right) &=& \pm (1-l)
\Bigg[ r-\rho_0+ \frac{\rho^2_+}{\rho_{+}-\rho_{-}}
\ln \left|  \frac{r-\rho_{+} }{\rho_{0}-\rho_{+} }\right|\\
&& -
\frac{\rho^2_-}{\rho_{+}-\rho_{-}}
\ln \left|  \frac{r-\rho_{-}}{\rho_{0}-\rho_{-}}\right|
\Bigg]\,.
\end{eqnarray}
In Fig. \ref{f1}, we show 
the coordinate time, for a massless particle. Here, 
an observer located at $\rho_0$ will measure an infinite coordinate time for the photon to reach the event horizon. However, when the test particles move in the opposite direction, they require an infinite
coordinate time to reach infinity.

\begin{figure}[H]
	\begin{center}
		\includegraphics[width=6.2 cm]{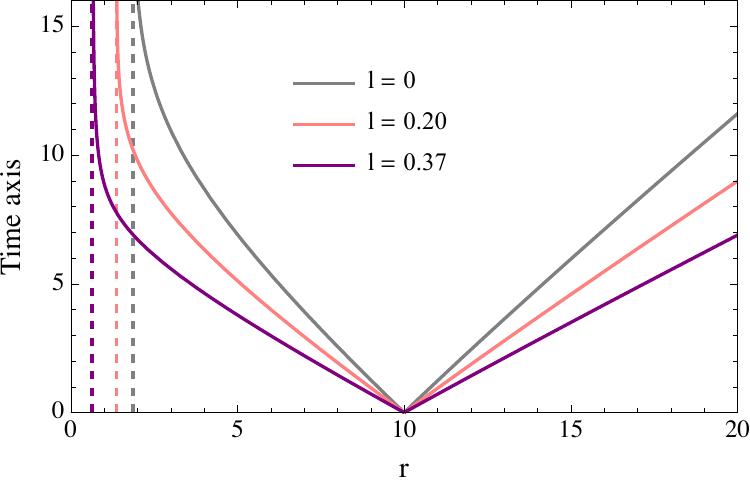}
	\end{center}
	\caption{
Radial motion of massless particles in the asymptotically flat KR black hole for different values of the Lorentz--violating parameter $l$. We have fixed $M=1$, $\Lambda=0$, $Q=0.5$ and $\rho_0=10$. The vertical dashed lines indicate the corresponding event horizons.}
	\label{f1}
\end{figure}

It is important to emphasize the role played by the Lorentz--violating parameter $l$ in Fig.~\ref{f1}. Although the radial motion with respect to the affine parameter is independent of the metric function, the coordinate time is sensitive to the KR background through the location of the horizons and through the overall factor $(1-l)$ in Eq.~(37). For the values considered, increasing $l$ shifts the event horizon $\rho_+$ towards smaller radial distances. Consequently, the logarithmic divergence of the coordinate time occurs closer to the event horizon. Away from the immediate vicinity of the horizon, the curves show that larger values of $l$ reduce the coordinate time required for the photon to move between a fixed radial position and the initial point $\rho_0$.

%\newpage
 % \textcolor{blue}{
	\item  Kalb--Ramond--de Sitter case $(\Lambda>0)$: 

The expression for the coordinate time is 
	 \begin{equation}\label{nb5}
	 t_{\Lambda}\left( r\right) =\pm{3(1-l)\over \Lambda}\sum_{i=1}^{4} F_i(r)\,,\\
	 \end{equation}%
	 where
     \begin{widetext}
	 \begin{eqnarray}\label{Fi}
	F_1(r)&=&{-R_{++}^2\over (R_{++}-R_{-})(R_{++}-R_{4})(R_{++}-R_{+})}\ln \left|  \frac{R_{++}-r}{R_{++}-\rho_{0}}\right|\,,\\ 
	F_2(r)&=&{R_{+}^2\over (R_{+}-R_{-})(R_{+}-R_{4})(R_{++}-R_{+})}\ln \left|  \frac{r-R_{+}}{\rho_{0}-R_{+}}\right| \,,\\ 
	F_3(r)&=&{-R_{-}^2\over (R_{-}-R_{4})(R_{++}-R_{-})(R_{+}-R_{-})}\ln \left|  \frac{r-R_{-}}{\rho_{0}-R_{-}}\right|\,,\\ 
	F_4(r)&=&{R_{4}^2\over (R_{-}-R_{4})(R_{++}-R_{4})(R_{+}-R_{4})}\ln \left|  \frac{r-R_{4}}{\rho_{0}-R_{4}}\right|\,.
	 \end{eqnarray}%
\end{widetext}
In Fig. \ref{f22}  we show the coordinate time, for a massless particle for the three cases considered for a fixed value of $M$, $Q$ and $\Lambda$. The effect of the Lorentz--violating parameter $l$ in Fig.~\ref{f22} is twofold. First, increasing $l$ modifies the location of the horizons: the event horizon $R_+$ is displaced toward smaller radial values, while the cosmological horizon $R_{++}$ is shifted outward. As a consequence, the static region available for photon propagation between $R_+$ and $R_{++}$ becomes wider. Second, the coordinate-time scale is affected by the explicit factor $(1-l)$ in Eq.~(38), as well as by the $l$-dependence of the residues associated with the horizon poles. Away from the immediate vicinity of the horizons, larger values of $l$ reduce the coordinate time required for the photon to move from the initial position $\rho_0$ to a fixed radial point. This is reflected in the fact that the curves with larger $l$ lie below the corresponding curve with $l=0$ in the bulk of the static region. Nevertheless, the KR parameter does not remove the coordinate-time divergences: an infalling photon still requires an infinite coordinate time to reach the event horizon, while an outgoing photon requires an infinite coordinate time to reach the cosmological horizon. Therefore, in the de Sitter case the Lorentz--violating parameter changes the size of the static patch and the coordinate-time scale of the radial motion, but preserves the qualitative causal behavior imposed by the two horizons. \\

\begin{figure}[H]
	\begin{center}
		\includegraphics[width=60mm]{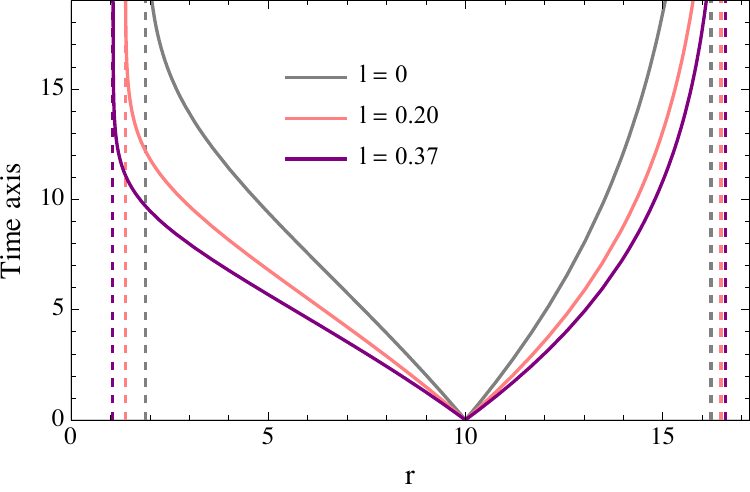}
	\end{center}
    \caption{
Radial motion of massless particles in the KR--de Sitter black hole for different values of the Lorentz--violating parameter $l$. We have fixed $M=1$, $Q=0.5$, $\Lambda=0.01$ and $\rho_0=10$. The vertical dashed lines indicate the event and cosmological horizons for each value of $l$.}
	\label{f22}
\end{figure}

\item  Kalb--Ramond-Anti-de Sitter $(\Lambda<0)$: 

In this case the coordinate time is given by 
\begin{equation}\label{nb6}
t_{\ell}\left( r\right) =\pm(1-l)\,{\ell^2}\sum_{i=1}^{4} \left( G_i(r)- G_i(\rho_{0})\right)\,,\ 
\end{equation}%
where
\begin{widetext}
\begin{eqnarray}\label{Gi}
G_1(r)&=&\frac{{r_+}^2 \ln (r-{r_+})}{({r_+}-{r_-}) ({r_3} {r_4}+{r_+} ({r_+}-{r_3}-{r_4}))}\,,\\ 
G_2(r)&=&-\frac{{r_-}^2 \ln (r-{r_-})}{({r_+}-{r_-}) ({r_3} {r_4}+{r_-} ({r_-}-{r_3}-{r_4}))}\,,\\ 
G_3(r)&=&-\frac{ ({r_3} {r_4} ({r_-}+{r_+})-{r_-} {r_+} ({r_3}+{r_4}))\ln \left(r^2-r ({r_3}+{r_4})+{r_3} {r_4}\right)}{2 ({r_3}{r_4}+{r_-} ({r_-}-{r_3}-{r_4})) ({r_3} {r_4}+{r_+} ({r_+}-{r_3}-{r_4}))}\,,\\ 
\notag G_4(r)&=&\frac{\left(2 {r_3}^2 {r_4}^2-{r_3} {r_4} (({r_-}+{r_+}) ({r_3}+{r_4})+2 {r_-} {r_+})+{r_-} {r_+} ({r_3}+{r_4})^2\right)}{\sqrt{4 {r_3} {r_4}-({r_3}+{r_4})^2} ({r_3} {r_4}+{r_-} ({r_-}-{r_3}-{r_4})) ({r_3} {r_4}+{r_+} ({r_+}-{r_3}-{r_4}))}\times\\
&\times &\tan ^{-1}\left(\frac{2 r-{r_3}-{r_4}}{\sqrt{4 {r_3} {r_4}-({r_3}+{r_4})^2}}\right) \,.
\end{eqnarray}%
\end{widetext}
In Fig.~\ref{f23}, the effect of the Lorentz--violating parameter $l$ is encoded both in the position of the event horizon and in the coordinate-time scale. For fixed $M$, $Q$ and AdS radius $\ell$, increasing $l$ shifts the outer horizon $r_+$ toward smaller radial values. Thus, the logarithmic divergence of the coordinate time occurs closer to the event horizon, enlarging the exterior radial interval displayed in the plot. In contrast with the de Sitter case, there is no cosmological horizon; instead, the asymptotically AdS behavior makes the coordinate time approach a finite limiting value as photons move toward large radial distances. This feature follows from the large-$r$ behavior of the metric function, for which $f(r)$ grows quadratically with $r$. Since the KR parameter rescales this asymptotic term through the factor $(1-l)^{-1}$, larger values of $l$ increase the effective growth of $f(r)$ and therefore reduce the coordinate time needed for photons to move away from the initial position $\rho_0$. Hence, the KR parameter does not change the qualitative AdS radial behavior, but it shifts the horizon inward and lowers the coordinate-time scale associated with photon propagation in the exterior region.
\begin{figure}[H]
	\begin{center}
		\includegraphics[width=70mm]{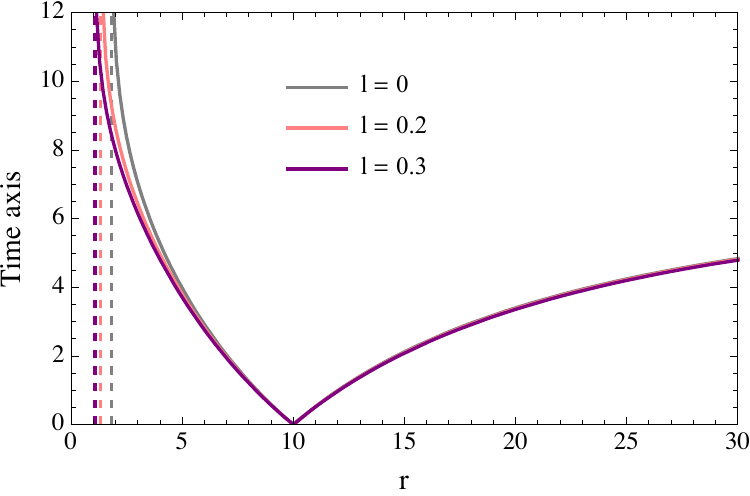}
	\end{center}
	\caption{
    Radial motion of massless particles in the KR--anti-de Sitter black hole for different values of the Lorentz-violating parameter $l$. We have fixed $M=1$, $Q=0.5$, $\ell=10$ ($\Lambda=-\frac{3}{\ell^{2}}$), and $\rho_0=10$. The vertical dashed lines indicate the corresponding event horizons. 
    }
	\label{f23}
\end{figure}

\end{enumerate}

Note that for the asymptotically flat case, as we mentioned, when the test particles move to the infinity, they require an infinite
coordinate time to reach infinity. On the other hand, when the spacetime is asymptotically dS  the test particles require an infinite coordinate time in order to reach the cosmological horizon $R_{++}$.\\ 
%So, the effect of the cosmological constant is to introduce a distance limit for $\Lambda>0$, and a time limit for $\Lambda<0$.\\

The asymptotic behavior of the coordinate time near horizons for different values of $\Lambda$ is described below

\begin{itemize}

\item For $ \Lambda = 0 $, we choose the negative sign in the solution (Eq.~34) to describe infall toward the event horizon. Then, the coordinate time near the event horizon is given by
\begin{equation}\label{t0}
t(r) \approx \frac{(1-l)\,\rho_+^2}{\rho_+ - \rho_2}
\ln \left| \frac{\rho_0 - \rho_+}{r - \rho_+} \right|\,,
\end{equation}
from which it is clear that as $ r \rightarrow \rho_+ $, we have $ t \rightarrow \infty $.

\item  For $ \Lambda > 0 $, both an event horizon and a cosmological horizon exist. The infall toward the event horizon, described by Eq.~(35), can be written near the event horizon as
$
t(r) \approx -\frac{3}{\Lambda} F_2(r),
$
which yields
\begin{equation}\label{t1a}
t(r) \approx
 \frac{3(1-l)\,R_+^2}{\Lambda(R_+ - R_3)(R_+ - R_4)(R_{++} - R_+)}
\ln \left| \frac{\rho_0 - R_+}{r - R_+} \right|\,,
\end{equation}
and again, as $ r \rightarrow R_+ $, it follows that $ t \rightarrow \infty $. Conversely, for the outgoing motion toward the cosmological horizon, also described by Eq.~(35), we have near the cosmological horizon
$
t(r) \approx \frac{3}{\Lambda} F_1(r)\,,
$
leading to
\begin{equation}\label{t1b}
t(r) \approx
\frac{3(1-l)\,R_{++}^2}{\Lambda(R_{++} - R_3)(R_{++} - R_4)(R_{++} - R_+)}
\ln \left| \frac{R_{++} - \rho_0}{R_{++} - r} \right|\,,
\end{equation}
so that as $ r \rightarrow R_{++} $, the coordinate time diverges: $ t \rightarrow \infty $.

\item  For $ \Lambda < 0 $, we again choose the negative sign in the solution (Eq.~40) to describe infall toward the event horizon. In this case, the coordinate time near the event horizon is given by
$
t(r) \approx -\ell^2 \left( G_1(r) - G_1(\rho_0) \right),
$
which leads to
\begin{equation}\label{t2}
t(r) \approx 
\frac{\ell^2\,(1-l)\,r_+^2}{(r_+ - r_2)(r_3 r_4 + r_+ (r_+ - r_3 - r_4))}
\ln \left| \frac{\rho_0 - r_+}{r - r_+} \right|\,,
\end{equation}
indicating that as $ r \rightarrow r_+ $, the coordinate time diverges: $ t \rightarrow \infty $. Furthermore, since $ r_4 = r_3^* $, it follows that $ r_4 + r_3 \in \mathbb{R} $ and $ r_4 r_3 \in \mathbb{R} $, ensuring that the coordinate time remains real.

\end{itemize}

\subsection{Angular motion}

\begin{figure}[!h]
	\begin{center}
		\includegraphics[width=60mm]{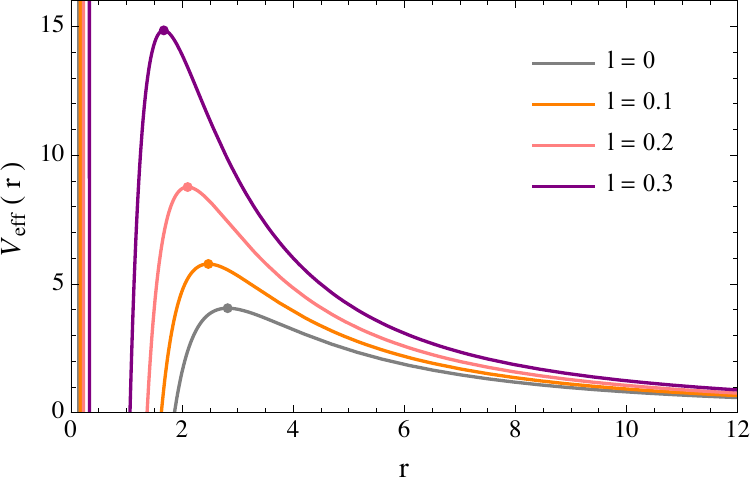}
        \includegraphics[width=60mm]{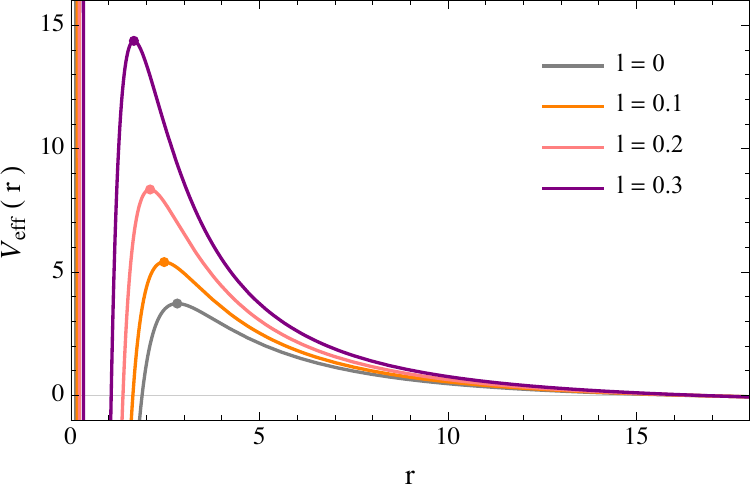}
        \includegraphics[width=60mm]{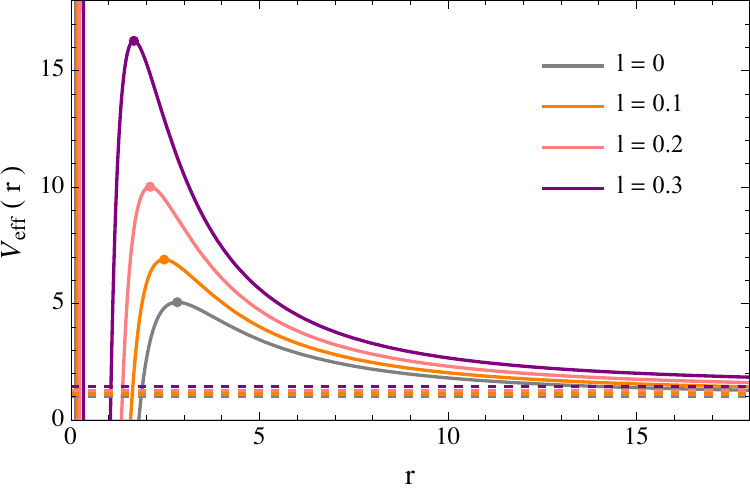}
	\end{center}
	\caption{Effective potential for null geodesics in the charged KR black hole spacetime for different values of the Lorentz--violating parameter $l$. 
The upper, middle and lower panels correspond respectively to the asymptotically flat, de Sitter ($\Lambda=0.01$), and anti-de Sitter cases ($\Lambda=-0.01$), respectively. 
We have fixed $M=1$, $Q=0.5$, and $L=10$. 
}
	\label{f4}
\end{figure}

%\textcolor{red}{
Now, we study the motion for $L\neq 0$ with the effective potential $V_{\text{eff}}$ from (\ref{g12}) in the case $m=0$, namely
\begin{equation}\label{n1}
V_{\text{eff}}\left( r\right) =\frac{L^{2}}{(1-l)\,r^{2}}-\frac{2ML^{2}}{r^{3}}+\frac{Q^{2}L^{2}}{(1-l)^2\,r^{4}}-
\frac{\Lambda L^{2}}{3(1-l)}\,.
\end{equation}
A typical graph of this effective potential is shown in Fig. \ref{f4}. Now, in order to calculate the value of $r$, where the effective potential is maximum ($r_u$), we consider $V'_{\text{eff}}(r)$ given by 
\begin{equation}\label{Vp}
V'_{\text{eff}}\left( r\right) =-\frac{2\,L^{2}}{(1-l)\,r^{3}}+\frac{6ML^{2}}{r^{4}}-\frac{4Q ^{2}L^{2}}{(1-l)^2\,r^{5}}\,,
\end{equation}
and the condition $V'_{\text{eff}}(r) = 0 \quad\Rightarrow \quad  r^2-3(1-l)Mr+{2Q^2\over 1-l}=0\,,
$
that corresponds to a quadratic equation, whose positive solution is
\begin{equation}
r_{u}={3M(1-l)\over 2}+\sqrt {{9M^2(1-l)^2\over 4}-{2\,Q ^2\over (1-l)} }\,,
\label{rul}
\end{equation}
which represents an unstable circular orbit. 
%}

 As shown in Fig.~\ref{f4}, the Lorentz--violating parameter $l$ has a direct effect on the shape of the effective potential. 
For fixed $M$, $Q$ and $L$, increasing $l$ raises the height of the potential barrier and shifts its maximum toward smaller radial distances. 
This behavior follows from the explicit $l$-dependence of the null effective potential, where the asymptotic contribution is weighted by $(1-l)^{-1}$ and the charge contribution by $(1-l)^{-2}$. 
Consequently, the KR background makes the photon sphere barrier more pronounced and modifies the characteristic scale at which photons can remain in unstable circular motion.

On the other hand, the position of the maximum, $r_u$, is independent of the cosmological constant. 
This can be understood from the structure of $V_{\rm eff}(r)$: for null geodesics, the term proportional to $\Lambda$ contributes as a constant shift to the effective potential and therefore changes its value but not the extremum condition. 
Thus, the cosmological constant modifies the energy threshold for the existence of unstable circular photon orbits, but it does not shift their radial location. 
Evaluating the effective potential Eq.~(\ref{n1}) at $r_u$, given by Eq.~(\ref{rul}), we obtain
$E_{u}^{\ell}>E_{u}^{0}>E_{u}^{\Lambda}$, and therefore
$b_{u}^{\ell}<b_{u}^{0}<b_{u}^{\Lambda}$. 
Hence, photons require a larger energy, or equivalently a smaller critical impact parameter, to remain on the unstable circular orbit in the AdS branch, whereas in the dS branch the required energy is smaller and the corresponding critical impact parameter is larger. 
The asymptotically flat case lies between these two regimes. 
Therefore, the KR parameter controls the local height and position of the photon sphere barrier, while the cosmological constant fixes the energetic scale associated with critical photon motion.

\begin{figure}[H]
	\begin{center}
		\includegraphics[width=60mm]{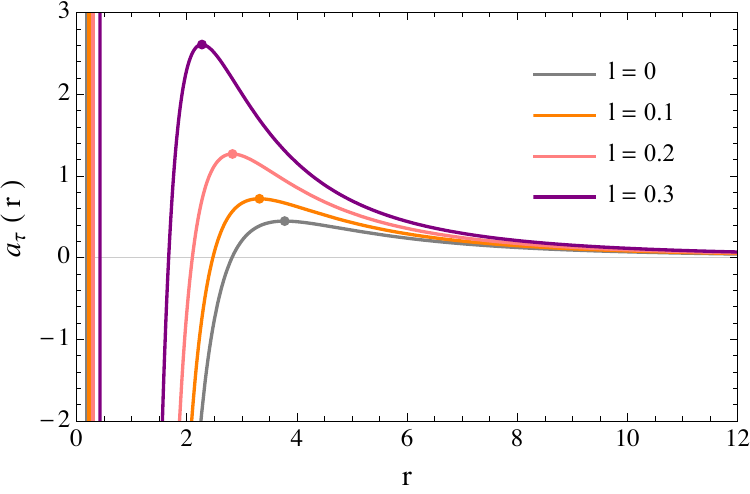}
	\end{center}
	\caption{Plot of the radial acceleration for massless particles. Here, $M=1$, $Q=0.5$, and $L=10$.
	The graph shows the radial acceleration, which is independent of the cosmological constant. The radial acceleration is maximum $a_{\lambda}(r_I)=0.447, 0.720, 1.268, 2.608$, at the inflection point, $r_I=3.780, 3.321, 2.832, 2.277$, for $l=0, 0.1, 0.2, 0.3$, respectively.}
	\label{f2b}
\end{figure}

%\textcolor{red}{
On the other hand, in Fig. \ref{f2b} we plot the radial acceleration for the massless particles,  defined as $ a_{r} \equiv \ddot{r} =-V_{\text{eff}}^{\prime}(r)/2$. From Eq.~(\ref{n1}) one obtains
\begin{equation}
a_r(r)=
\frac{L^2}{(1-l)r^3}
-\frac{3ML^2}{r^4}
+\frac{2Q^2L^2}{(1-l)^2r^5}\,.
\end{equation}
Notice that $a_r$ is independent of the cosmological constant. This follows from the fact that, for null geodesics, the $\Lambda$-term in $V_{\rm eff}$ contributes only as a constant shift and therefore disappears after taking the radial derivative. In contrast, the KR parameter $l$ remains explicitly in the radial acceleration through the factors $(1-l)^{-1}$ and $(1-l)^{-2}$. Now, in order to calculate the value of $r$, where the radial acceleration is maximum ($r_I$), we consider $V''_{\text{eff}}(r)$ given by 
\begin{equation}\label{Vpp}
V''_{\text{eff}}\left( r\right) =\frac{6\,L^{2}}{(1-l)\,r^{4}}-\frac{24ML^{2}}{r^{5}}+\frac{20Q ^{2}L^{2}}{(1-l)^2\,r^{6}}\,,
\end{equation}
and the condition $V''_{\text{eff}}(r) = 0 \quad\Rightarrow \quad  r^2-4(1-l)Mr+{10Q^2\over 3(1-l)}=0$, which corresponds to a quadratic equation whose positive solution is
\begin{equation}
r_{I}=2M(1-l)+\sqrt {4M^2(1-l)^2-{10\,Q ^2\over 3(1-l)} }\,.
\label{n8l}
\end{equation}

This radius corresponds to the inflection point of the effective potential and to the maximum of the positive radial acceleration outside the unstable circular photon orbit. Fig.~\ref{f2b} shows that the Lorentz--violating parameter has a significant effect on the radial acceleration. As $l$ increases, both the zero of $a_r$, located at the unstable circular orbit $r_u$, and the maximum located at $r_I$, are shifted toward smaller radial distances. At the same time, the maximum value of $a_r$ increases. Therefore, the KR background makes the effective acceleration profile steeper and enhances the outward radial acceleration in the region outside the photon sphere. This behavior is consistent with the increase of the height of the effective-potential barrier observed in Fig.~\ref{f4}.

The sign of $a_r$ also provides a useful interpretation of the photon dynamics. In the region between the event horizon and the unstable circular orbit, $r_+<r<r_u$, the radial acceleration is negative, indicating that photons are driven toward the black hole. At $r=r_u$, the radial acceleration vanishes, corresponding to the unstable circular null orbit. For $r_u<r<\infty$, the radial acceleration becomes positive and reaches its maximum at $r=r_I$, before decreasing asymptotically to zero at large distances. Thus, the KR parameter does not change the qualitative sign structure of the radial acceleration, but it shifts the characteristic radii inward and increases the strength of the acceleration outside the photon sphere.

%\newpage

%\textcolor{red}{
Now, in order to perform a qualitative analysis of the effective potential, we will consider the impact parameter. So, considering Eq. (\ref{g11}), and (\ref{n1}), we obtain 
\begin{equation}\label{em1}
\left(\frac{dr}{d\phi}\right)^{2}=\left(\frac{1}{b^2}+ \frac{\Lambda }{3(1-l)}\right)r^{4} -{r^{2}\over 1-l}+2Mr-{Q^{2}\over (1-l)^2}\,,
\end{equation}%
where $b \equiv L/E$ is the impact parameter.
%}
Therefore, based on the impact parameter values and Fig.~\ref{f4}, we present a brief qualitative description of the allowed angular motions for photons in the charged KR black hole spacetime. 

\begin{itemize}
	\item[(i)] {\it Capture zone:} 	If $0 < b < b_u$, the photons fall inexorably to the horizon $r_+$, or escape to infinity, depending on the initial conditions, and their cross section, $\sigma$, in this geometry is $\sigma=\pi\,b_u^2={\pi\,r_u^2 \over f(r_u)}$, which can be written explicitly as
    
	\begin{equation}
	\sigma={\pi\,(1-l)^2\,r_u^4 \over Q^2-2(1-l)^2Mr_u+(1-l)\,r_u^2-{\Lambda (1-l)\over 3}\,r_u^4}\,.
	\label{sigma}
	\end{equation}
   
	\item [(ii)] {\it Critical trajectories:} 
	If $b=b_u$, photons can stay in one of the unstable inner circular orbits of radius $r_u$. Therefore, photons that arrive from the initial distance $r_i$ ($r_+ < r_i < r_u$, or $r_u < r_i < \infty$) can asymptotically fall to a circle of radius $r_u$. The affine period in such orbit is
		\begin{equation}
	T_{\lambda}={2\pi\,r_u^2 \over L}\,,
	\label{T1}
	\end{equation}
		and the coordinate period is $T_{t}=2\pi\,b_u={2\pi\,r_u \over \sqrt{f(r_u)}}$, or, more explicitly,
          		\begin{equation}
		T_{t}={2\pi\,(1-l)\,r_u^2 \over \sqrt{ Q^2-2(1-l)^2Mr_u+(1-l)\,r_u^2-{\Lambda (1-l)\over 3}\,r_u^4}}\,.
		\label{T2}
		\end{equation}
      
	\item [(iii)] {\it Deflection zone:} For $\Lambda=0$ or $\Lambda>0$, the deflection zone is characterized by impact parameters satisfying $b_u < b < \infty$. On the other hand, if $\Lambda<0$, and $b_u < b < b_{\ell}$, this zone presents orbits of the first and second kind. The orbits of the first kind are allowed in the interval $r_d \leq r < \infty$, where the photons can come from a finite distance or from  infinity  until they reach the distance $r = r_d$ (which is a solution of the equation $V_{\text{eff}}(r_d) = E^2$), and then the photons are deflected. Note that photons with $b \geq b_{\ell}= {\ell}\sqrt{1-l}$ are not allowed in this zone. The orbits of the second kind are allowed in the interval $r_+ < r \leq r_ F$, where the photons come from a distance greater than the event horizon, then they reach the distance $r_F$ (which is a solution of the equation $V_{\text{eff}}(r_F) = E^2$) and then they plunge into the horizon.	
		\item [(iv)] {\it Second kind and lima\c{c}on of Pascal with precession} 
		If $b_u \leq b < \infty$, the return point is in the range $r_+ < r < r_u$, and then the photons plunge into the horizon. However, when $b = b_{\ell}$ a special geodesic can be obtained, known as the lima\c{c}on of Pascal with precession.	
	\end{itemize}

\subsubsection{First kind trajectories}

Now, in order to obtain the bending of light, we consider
Eq. (\ref{em1}), which can be written as
\begin{equation}	
\label{em}
\left(\frac{dr}{d\phi}\right)^{2}
\equiv {\mathcal{P}(r)\over \mathcal{B}^2}\,,
\end{equation}
where
\begin{eqnarray}
    \mathcal{P}(r) =  r^{4}- {
    \mathcal{B}^2\over (1-l)}r^{2}+2M\mathcal{B}^2r-{
    Q^2\mathcal{B}^2\over (1-l)^2}\,,
\end{eqnarray}
and
\begin{eqnarray}
    \mathcal{B} = \left(\frac{1}{b^2}+ \frac{\Lambda }{3\,(1-l)}\right)^{-1/2}\,,
\end{eqnarray}
is the {\it anomalous impact parameter},  defined as the combination of the usual impact parameter $ 1/b^2 $ and the cosmological constant $ \Lambda $.

\begin{itemize}
\item  For $ \Lambda > 0 $, we always have $ \frac{1}{\mathcal{B}^2} > 0 $, and the null geodesics are qualitatively similar to those in the Schwarzschild case.

\item For $ \Lambda < 0 $, the expression becomes
\begin{equation}
\frac{1}{\mathcal{B}^2} = \frac{1}{b^2} - \frac{1}{(1-l)\,\ell^2}\,,
\end{equation}
where $ \ell $ is the AdS radius. This formulation is particularly relevant in AdS spacetimes, as it allows for the characterization of a new region of spacetime that is not permitted in the $ \Lambda \geq 0 $ cases. When $ \frac{1}{\mathcal{B}^2} \leq 0 $, new second class trajectories emerge, associated with energies in the range $ 0 < E \leq E_\ell $. In the special case where $ \frac{1}{\mathcal{B}^2} = 0 $, the second-class trajectory corresponds to a limaçon of Pascal with precession. This new region, along with the appearance of exotic geodesics, is a distinctive feature of spherically symmetric AdS spacetimes \cite{Villanueva:2013zta,Gonzalez:2020zfd}.
\end{itemize}

Now, in order to obtain the return points, we solve the equation $\mathcal{P}(r ) = 0$,  so, the turning point is located at%
\begin{eqnarray}
    r_{d}&=&\alpha +\sqrt{{\mathcal{B}^2 \over 2(1-l)}-\alpha^2-{M\mathcal{B}^2 \over 2\,\alpha}}\,,\\
    r_{f}&=&\alpha -\sqrt{{\mathcal{B}^2 \over 2(1-l)}-\alpha^2-{M\mathcal{B}^2 \over 2\,\alpha}}\,,\\
    r_{3}&=&-\alpha +\sqrt{{\mathcal{B}^2 \over 2(1-l)}-\alpha^2+{M\mathcal{B}^2 \over 2\,\alpha}}\,,\\
    r_{4}&=&-\alpha -\sqrt{{\mathcal{B}^2 \over 2(1-l)}-\alpha^2+{M\mathcal{B}^2 \over 2\,\alpha}}\,,
\end{eqnarray}
where
\begin{eqnarray}
    \alpha &=&\sqrt{U_0\cosh \left[ \frac{1}{3}\cosh^{-1} \Xi_0\right]+\frac{\mathcal{B}^2}{6(1-l)}}\,,\\
    U_0&=&{\mathcal{B} \sqrt{\mathcal{B}^2-12Q^2} \over 6(1-l)}\,, \\
    \Xi_0&=&\frac{54M^2\mathcal{B}(1-l)^3-36Q^2\mathcal{B}-\mathcal{B}^3}{(\mathcal{B}^2-12Q^2)^{3/2}}\,,
\end{eqnarray}
$ r_d$ is the deflection distance, $ r_f$ is the return point and $ r_3$ with $ r_4$ are negative solutions. 

Now, using Eq. (\ref{em}), we obtain the quadrature
\begin{equation}\label{n9}
\phi \left( r\right) =\mathcal{B}\int_{r_{d}}^{r}\frac{\,dr'}{\sqrt{\left( r'-r_{d}\right)
		\left( r'-r_{f}\right) \left( r'-r_{3}\right) \left( r'-r_{4}\right) }}\,.
\end{equation}%

In order to integrate out (\ref{n9}) it is instructive
to make the change of variable $r=r_{d}\left( 1+{1\over 4U-\alpha_d/3}\right) $, and
after a brief manipulation, we obtain the polar
trajectory of massless particles
\begin{equation}\label{n11}
r\left( \phi \right) =r_{d}+{r_{d}\over 4\wp \left(\kappa_d\,\phi
	;g_{2},g_{3}\right)-\alpha_d/3}\,,
\end{equation}%
where 
\begin{eqnarray}
    \kappa_{d}&=&{r_d\over \mathcal{B}\sqrt{u_4u_3u_f}}\,, \\
    \alpha_{d}&=&u_4+u_3+u_f\,,
\end{eqnarray}
while $\wp \equiv \wp(y; g_2, g_3)$ is the $\wp$-Weierstra{\ss} function and
$g_{2}$ with $g_{3}$ are the so-called
Weierstra{\ss} invariants, given by

\begin{eqnarray}
\notag  g_{2}&=&\frac{1}{12}\left(  u_{4}^2+u_{3}^2+u_{f}^2-u_{4}\,u_{3}-u_{4}\,u_{f}-u_{f}\,u_{3}\right)\,,\\
\notag g_{3}&=&\frac{1}{432}\left(2 u_{4}-u_{3}-u_{f}  \right) \left(2 u_{3}-u_{4}-u_{f}  \right)\cdot \\
&&\left( u_{4}+u_{3}-2u_{f}  \right) \,.
\end{eqnarray}

\begin{figure}[!h]
	\begin{center}
		\includegraphics[width=55mm]{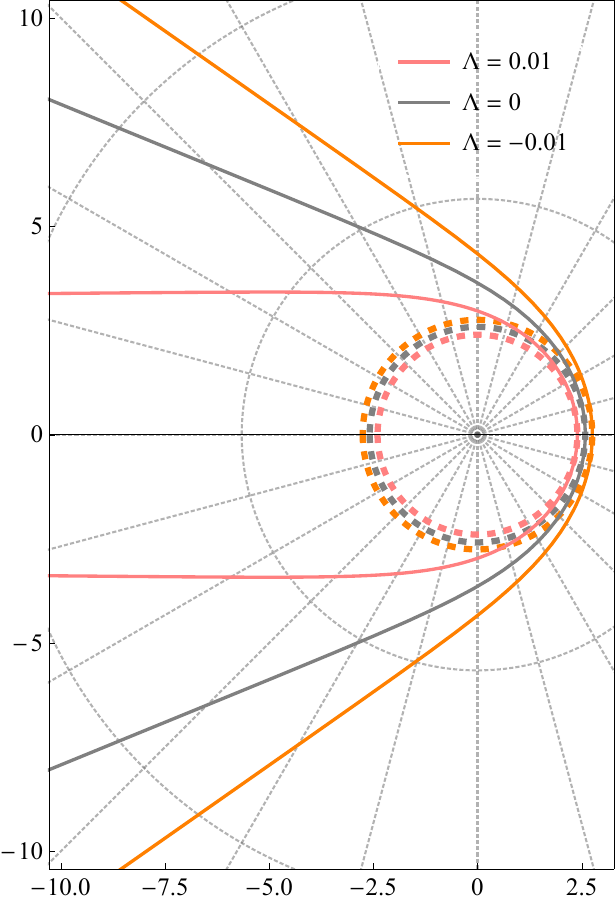}
	\end{center}
	\caption{Polar plot of the deflection trajectories of massless particles in the charged KR black hole spacetime. 
The three curves correspond to the de Sitter $(\Lambda>0)$, asymptotically flat $(\Lambda=0)$ and anti-de Sitter $(\Lambda<0)$ cases. 
We have fixed $M=1$, $Q=0.5$, $l=0.2$ and $L=10$, and all trajectories have the same energy $E^2=3$.  
}
	\label{f3}
\end{figure}

The other constants are: $u_{4}={r_d\over r_d-r_4}$, $ u_{3}={r_d\over r_d-r_3}$ and $u_{f}={r_d\over r_d-r_f}$. Fig.~\ref{f3} shows the polar trajectories associated with the deflection of light for the three asymptotic sectors of the charged KR geometry. We have fixed the black hole parameters $M$, $Q$ and the Lorentz--violating parameter $l$, so that the differences between the curves are produced only by the cosmological constant. The de Sitter trajectory bends more strongly than the asymptotically flat one, while the anti-de Sitter trajectory presents a larger distance of closest approach. This behavior can be understood from the anomalous impact parameter $\mathcal{B}$, which contains the combination between the usual impact parameter and the cosmological constant. A positive cosmological constant increases the effective value of $1/\mathcal{B}^2$, allowing the photon to approach the black hole more closely and producing a larger bending angle. In contrast, for $\Lambda<0$ the AdS contribution reduces $1/\mathcal{B}^2$, increasing the turning point $r_d$ and opening the trajectory outward.

The KR parameter enters this behavior through the factors $(1-l)^{-1}$ and $(1-l)^{-2}$ in the radial polynomial $\mathcal{P}(r)$. Therefore, even though Fig.~\ref{f3} is drawn for a fixed value of $l$, the KR background modifies the turning points and the angular scale of the orbit with respect to the RN--(A)dS case. In particular, the Lorentz--violating parameter changes the balance between the mass, charge and cosmological contributions in the trajectory equation. Thus, the figure illustrates that the global deflection geometry is controlled by $\Lambda$, while the detailed location of the deflection point and the shape of the orbit are affected by the KR background.

\bigskip

It is well known that photons can escape to infinity during a deflection process. So, considering $r(\phi)|_{\phi=0} = r_d$, the shortest distance to the black hole at which the deflection happens, and assuming that the incident photons come from infinity and escape to infinity, we have $r(\phi)|_{\phi_{\infty}} = \infty$. Now, by using Eq. (\ref{n11}) we obtain that the deflection angle, $\hat{\alpha}=2|\phi_{\infty}|-\pi$, is given by
\begin{equation}\label{angdef}
\hat{\alpha}={2\over \kappa_d} \left| \wp^{-1}\left({\alpha_d\over 12} 	;g_{2},g_{3}\right) \right| -\pi\,.
\end{equation}

\begin{figure}[!h]
	\begin{center}
		\includegraphics[width=60 mm]{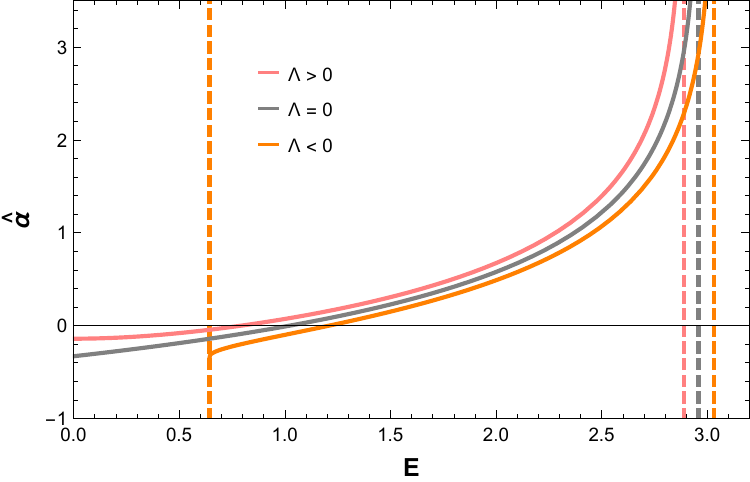}
	\end{center}
	\caption{The behavior of the deflection angle $\hat{\alpha}$ in terms of $E$, with  $M=1$, $Q=0.5$, $l=0.2$, and $L = 10$. $E_{u}^{\ell}=3.029$, $E_{u}^0=2.959$, $E_{u}^{\Lambda}=2.888$, $E_{\ell}=0.645$.}
	\label{defleccion2}
\end{figure}

The behavior of the deflection angle as a function of the photon energy is shown in Fig.~\ref{defleccion2}. 
For fixed values of the black hole parameters $M$, $Q$, the angular momentum $L$, and the Lorentz--violating parameter $l$, the three curves display the same qualitative feature: the deflection angle grows as the photon energy approaches the critical value $E_u$ associated with the unstable circular null orbit. 
In this limit the photon spends an arbitrarily long time in the vicinity of the photon sphere before escaping, and the deflection angle diverges. 
The position of this divergence depends on the asymptotic sector of the spacetime. 
Since the cosmological constant shifts the value of the effective potential at its maximum without changing the radius $r_u$, the critical energies satisfy
$E_u^{\ell}>E_u^{0}>E_u^{\Lambda}$, or equivalently
$b_u^{\ell}<b_u^{0}<b_u^{\Lambda}$. 
Therefore, for a fixed deflection angle, a photon requires a larger energy in the AdS branch than in the asymptotically flat or dS branches.

At low energies the three cases display a finite negative deflection angle. 
For the asymptotically flat and de Sitter branches, this behavior appears in the limit $E\rightarrow0$, where $\hat{\alpha}$ approaches a negative constant rather than vanishing. 
This negative value indicates an effective repulsive deflection, in the sense that the photon trajectory bends outward with respect to the reference direction. 
Such behavior should not be interpreted as a genuine repulsive gravitational force, but as a consequence of the modified null effective potential produced by the KR deformation.

In the AdS branch, deflection trajectories of the first kind exist only above the threshold energy $E_{\ell}$, determined by the condition $\mathcal{B}^{-2}>0$. 
As $E\rightarrow E_{\ell}^{+}$, the deflection angle also approaches a finite negative value. 
Then, as the photon energy increases, $\hat{\alpha}$ grows, changes sign, and eventually diverges when the energy approaches the critical value $E_u$ associated with the unstable circular null orbit. 
Therefore, the zero of $\hat{\alpha}$ marks the transition between an outward, or repulsive, deflection regime and the usual attractive bending regime. Thus, Fig.~\ref{defleccion2} shows that the cosmological constant controls the energetic domain and the critical threshold of the deflection trajectories, whereas the KR parameter enters through the modified photon-sphere scale and the corresponding critical impact parameter.

\bigskip

\begin{figure}[!h]
	\begin{center}
		\includegraphics[width=50mm]{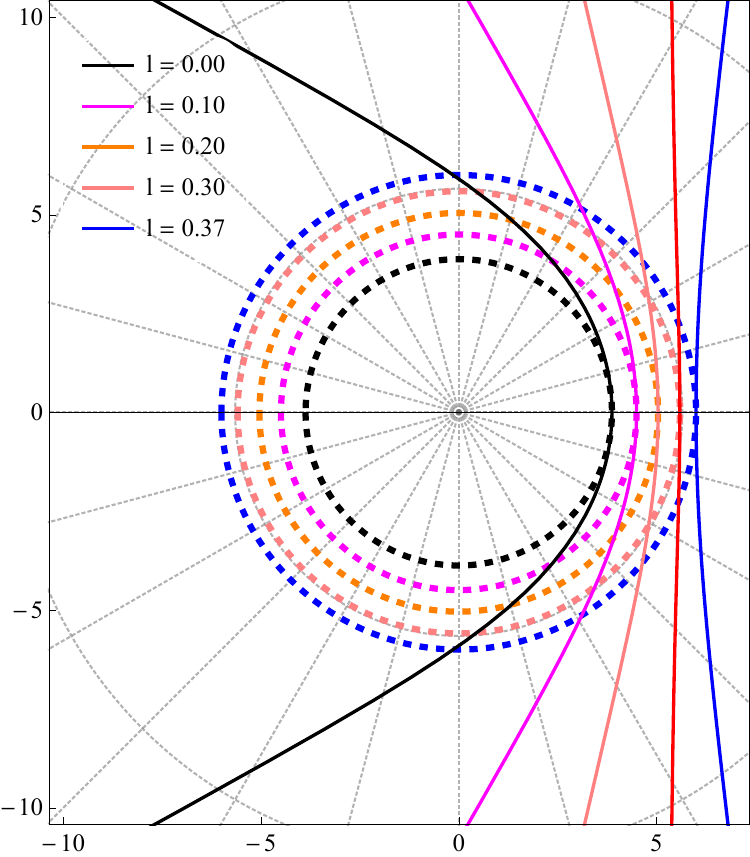}
        \includegraphics[width=50mm]{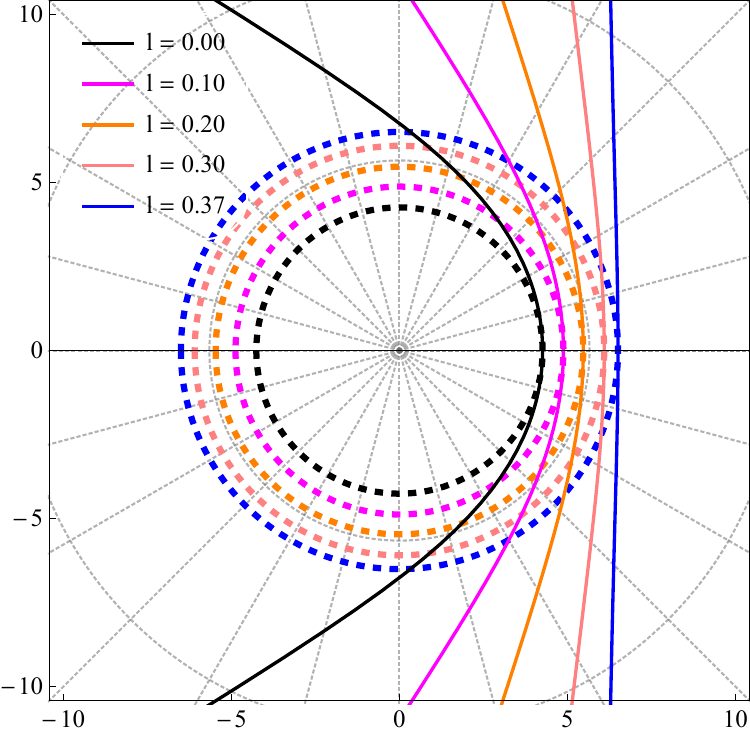}
        \includegraphics[width=50mm]{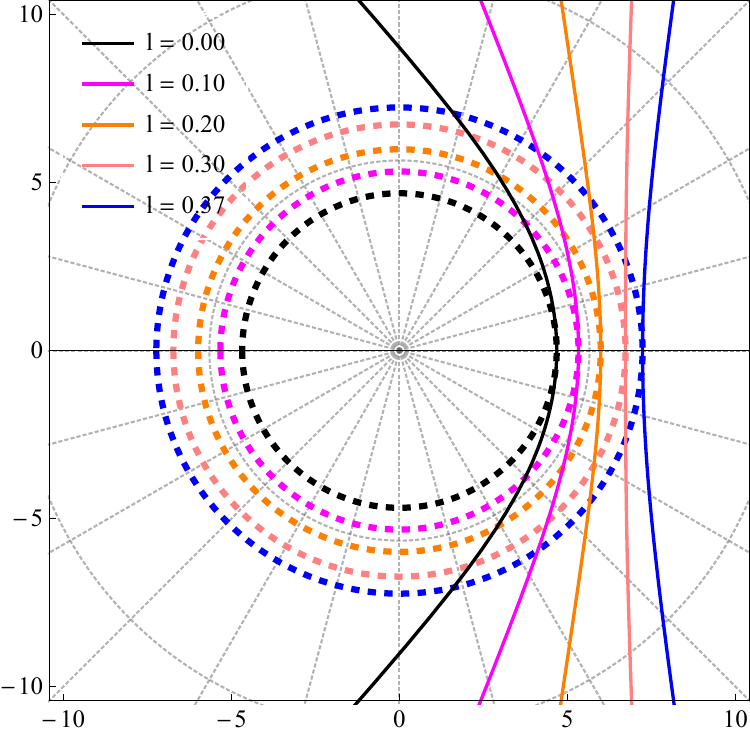}
	\end{center}
	\caption{Polar plot of the deflection trajectories of massless particles in the charged KR black hole spacetime. We have fixed $M=1$, $Q=0.5$, and $L=10$, and all trajectories have the same energy $E^2=3$. Top panel for $\Lambda=0.01$, central panel for $\Lambda =0$, and bottom panel for $\Lambda=-0.01$}
	\label{f33}
\end{figure}

Fig.~\ref{f33} shows the effect of the Lorentz--violating parameter $l$ on the first-kind trajectories associated with the bending of light. 
In each panel, the curve with $l=0$ represents the RN--(A)dS limit and will be used as the reference trajectory. 
Therefore, the remaining curves show how the KR background deforms the standard charged black-hole deflection geometry. 
We have fixed $M$, $Q$, $L$ and the photon energy, so that, for a given value of $\Lambda$, the differences among the trajectories are entirely due to the Lorentz-violating parameter. 

For small values of $l$, the trajectories remain close to the RN--(A)dS reference case and display the usual attractive bending: photons coming from the asymptotic region approach the black hole, reach an external turning point $r_d$, and then escape. 
However, as $l$ increases, the radial polynomial governing the angular motion is modified through the factors $(1-l)^{-1}$ and $(1-l)^{-2}$. 
These terms enhance the effective barrier felt by photons with fixed energy and angular momentum. 
As a consequence, the external turning point is displaced outward with respect to the $l=0$ case. 
Thus, the photon is reflected farther away from the black hole before probing the region where the mass term dominates the attractive bending.

For sufficiently large values of $l$, the deformation becomes strong enough to change the apparent character of the deflection trajectory. 
The curves then develop an outward-bending behavior when compared with the RN--(A)dS reference orbit. 
This should not be interpreted as a genuine repulsive gravitational force, but rather as an effective repulsive deflection induced by the KR modification of the null effective potential. 
In this regime, the enhanced barrier prevents the photon from approaching the black hole as closely as in the $l=0$ geometry, producing a trajectory that opens outward.

It is also important to distinguish this behavior from the shift of the photon-sphere radius $r_u$. 
Although $r_u$ moves inward as $l$ increases, the turning point of a non-critical deflection trajectory depends on the value of the photon energy relative to the effective-potential barrier. 
Since the KR parameter raises this barrier, a photon with fixed $E$ and $L$ reaches its turning point before approaching the photon sphere. 
Therefore, the bending trajectories become more external as $l$ grows, even though the circular null orbit itself is displaced inward. 
This shows that the KR background affects not only the critical circular orbit, but also the non-critical deflection trajectories responsible for the deflection of light.

The comparison between the three panels shows that the cosmological constant controls the global opening of the trajectories. 
For fixed $l$, the de Sitter branch allows photons to approach the black hole more closely, whereas the anti-de Sitter branch shifts the turning point outward. 
This is consistent with the role of the anomalous impact parameter $\mathcal{B}$: a positive cosmological constant increases $\mathcal{B}^{-2}$, while a negative cosmological constant decreases it. 
Hence, Fig.~\ref{f33} illustrates the combined effect of the KR Lorentz-violating parameter and the cosmological constant on the deflection sector of null geodesics, with the $l=0$ RN--(A)dS trajectory providing the natural benchmark for measuring the deformation.

\newpage

\subsubsection{Second kind trajectories and  lima\c{c}on of Pascal with precession}

The spacetime allows second kind trajectories, when $b_u < b<\infty$, where the turning point is in the range $r_+ <r <r_u$, and then the photons plunge into the horizon. 
In order to integrate out (\ref{n9}) from the initial distance $r_f$ to $r$, it is useful
to make the change of variable $r=r_{f}\left( 1-{1\over 4U-\alpha_f/3}\right) $, and
after a brief manipulation, we obtain the polar
trajectory of massless particles
\begin{equation}\label{n111}
r\left( \phi \right) =r_{f}-{r_{f}\over 4\wp \left(\kappa_f\,\phi
	;g_{2f},g_{3f}\right)-\alpha_f/3}\,,
\end{equation}%
where 
\begin{eqnarray}
    \kappa_{f}&=&\frac{r_f}{ \mathcal{B}\sqrt{u_{4f}u_{3f}u_d}} \\
    \alpha_{f}&=&u_d-u_{3f}-u_{4f}\,,
\end{eqnarray}
while $\wp \equiv \wp(y; g_{2f}, g_{3f})$ is the $\wp$-Weierstra{\ss} function and
$g_{2f}$ with $g_{3f}$ are the so-called
Weierstra{\ss} invariants, given by

\begin{eqnarray}
 \notag g_{2f}&=&\frac{1}{12}\left(  u_{4f}^2+u_{3f}^2+u_{d}^2-u_{4f}\,u_{3f}+u_{4f}\,u_{d}+u_{d}\,u_{3f}\right)\,,\\
\notag g_{3f}&=&\frac{1}{432}\left(u_{3f}-u_{d}-2 u_{4f}  \right) \left(2 u_{3f}-u_{4f}+u_{d}  \right)\cdot \\
&& \left( u_{4f}+u_{3f}+2u_{d}  \right)\,. 
\end{eqnarray}

\begin{figure}[!h]
	\begin{center}
		\includegraphics[width=70mm]{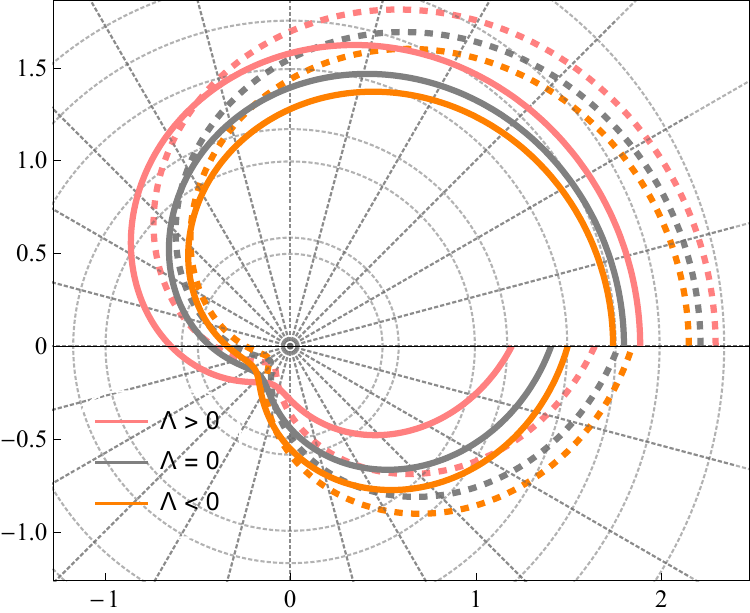}
	\end{center}
	\caption{Polar plot for second kind trajectories with $M=1$, $Q=0.5$, and $L = 10$. Solid curves for $l=0.2$ and $E^2=8.0$. Dashed curves for $l=0$ and $E^2=3.0$. The different energies are chosen so that the trajectories remain in the second-kind sector of each geometry.}
	\label{caida}
\end{figure}

The other constants are: $u_{4f}={r_f\over r_f-r_4}$, $ u_{3f}={r_f\over r_f-r_3}$ and $u_{d}={r_f\over r_d-r_f}$. Fig.~\ref{caida} shows representative second-kind null trajectories in the charged KR black-hole spacetime and compares them with the corresponding RN--(A)dS reference geometry. 
The dashed curves correspond to the $l=0$ case, where the RN--(A)dS geometry is recovered, while the solid curves correspond to the Lorentz-violating case $l=0.2$. 
Unlike Fig.~\ref{f33}, where first-kind deflection trajectories were compared at fixed photon energy, here the energies have been chosen differently in each geometry in order to keep the trajectories inside the second-kind sector. 
This is necessary because the KR parameter modifies the effective potential, the critical impact parameter and the allowed radial domain for plunging orbits.

Second-kind trajectories do not describe photons coming from infinity and escaping again to infinity. 
Instead, the photon moves in the inner allowed region, reaches a finite return point $r_f$, and then plunges into the event horizon. 
Thus, the relevant comparison is not between two photons with the same energy, but between two representative plunging trajectories belonging to the same dynamical class in two different geometries: the RN--(A)dS background and its KR deformation.

With this interpretation, the separation between the dashed and solid curves should be understood as the combined effect of the Lorentz-violating deformation and the corresponding shift of the energy range in which second-kind trajectories exist. 
Turning on $l$ changes the radial polynomial through the factors $(1-l)^{-1}$ and $(1-l)^{-2}$, modifying the balance between the mass, charge and angular-momentum terms in the effective potential. 
Consequently, the return point and the angular extension of the plunging trajectory are modified with respect to the RN--(A)dS case. 
The solid curves therefore illustrate how the KR background changes the inner plunging sector of null geodesics, rather than representing a fixed-energy deformation of the dashed RN--(A)dS trajectories.

The comparison among the three branches also shows the role of the cosmological constant. 
For the same class of second-kind motion, the de Sitter, asymptotically flat and anti-de Sitter sectors display different radial extensions and angular openings. 
This reflects the fact that $\Lambda$ modifies the global structure of the allowed region, while the KR parameter changes the local effective barrier and the location of the relevant turning points. 
Hence, Fig.~\ref{caida} should be interpreted as a comparison between dynamically equivalent second-kind trajectories in RN--(A)dS and KR-deformed geometries, with the energy adjusted in each case to remain in the appropriate plunging regime.

\bigskip

{\bf{Lima\c{c}on of Pascal with precession.}} 
The AdS spacetime allows second kind trajectories, when $b_{\ell} < b<\infty$, where the turning point is in the range $r_+ <r <r_0$, and then the photons plunge into the horizon. However, there is a special geodesic for AdS spacetime, which can be obtained when the anomalous impact parameter is $\mathcal{B }\rightarrow\infty \, (b = \ell\sqrt{1-l})$. In this case, the radial coordinate is restricted to $r_+ < r < r_0$, and the equation of motion (\ref{em1}) can be written as
\begin{equation}\label{lima1}
\phi(r)=-\sqrt{1-l}\int_{r_0}^{r} {dr'\over  \sqrt{-r'^{2}+2M(1-l)r'-{Q^2\over 1-l}}}  \,,        
\end{equation}
and the return points are
\begin{eqnarray}\label{lima2}
r_{0}&=&M(1-l)+\sqrt{M^2(1-l)^2-{Q^2\over 1-l}}\,,\\
d_{0}&=&M(1-l)-\sqrt{M^2(1-l)^2-{Q^2\over 1-l}}\,.
\end{eqnarray}
Thus, it is straightforward to find the solution of Eq. (\ref{lima1}), which is given by
\begin{equation}\label{lima3}
r(\phi)=M(1-l)+\sqrt{M^{2}(1-l)^2-{Q^2\over 1-l}}\,\cos {\phi \over \sqrt{1-l}} \,, 
\end{equation}%
which represents a Kalb--Ramond deformation of the Pascal lima\c{c}on-type geodesic, with a precession controlled by the Lorentz-violating parameter.
The curve with $l=0$ corresponds to the RN--AdS reference case, while the curves with $l>0$ describe the deformation induced by the KR background.

The main effect of increasing $l$ is twofold. First, the radial size of the orbit decreases. This can be seen from the displacement of the outer limiting radius $r_0$, which moves inward as the Lorentz--violating parameter grows. At the same time, the inner scale $d_0$ is shifted outward. Therefore, the radial domain available to the lima\c{c}on-type trajectory becomes narrower with respect to the RN--AdS case. Second, the angular dependence is modified through the factor $\phi/\sqrt{1-l}$, which produces a precession of the orbit. As a consequence, the KR deformation does not simply rescale the RN--AdS lima\c{c}on, but changes both its radial extension and its angular structure.

This behavior is consistent with the interpretation developed for the previous figures. The Lorentz--violating parameter enhances the effective barrier felt by photons and changes the balance between the mass, charge and angular-momentum contributions to the null geodesic equation. In the present limiting case, this produces a more compact and precessing lima\c{c}on when compared with the $l=0$ reference geometry. 

It is important to emphasize that this trajectory belongs to the inner null-geodesic structure. Unlike the first-kind deflection orbits associated with the usual bending of light, the lima\c{c}on-type curve is not an escape trajectory from infinity. Rather, it appears as a limiting second-kind orbit in the AdS geometry. Thus, its physical relevance is mainly theoretical: it completes the classification of null geodesics and provides a clear analytic signal of how the KR Lorentz--violating parameter deforms the RN--AdS geodesic structure. 

\bigskip
\begin{figure}[!h]
	\begin{center}
    \includegraphics[width=60mm]{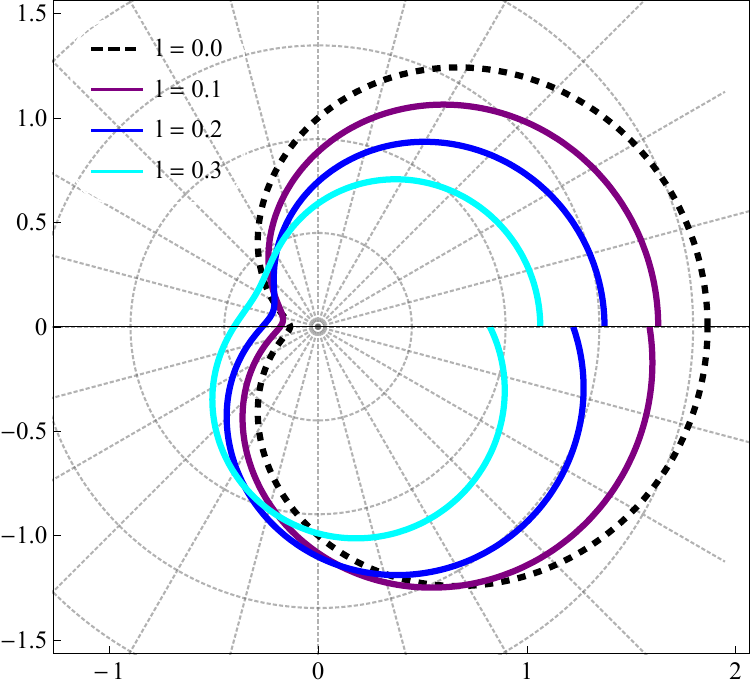}
		\includegraphics[width=60mm]{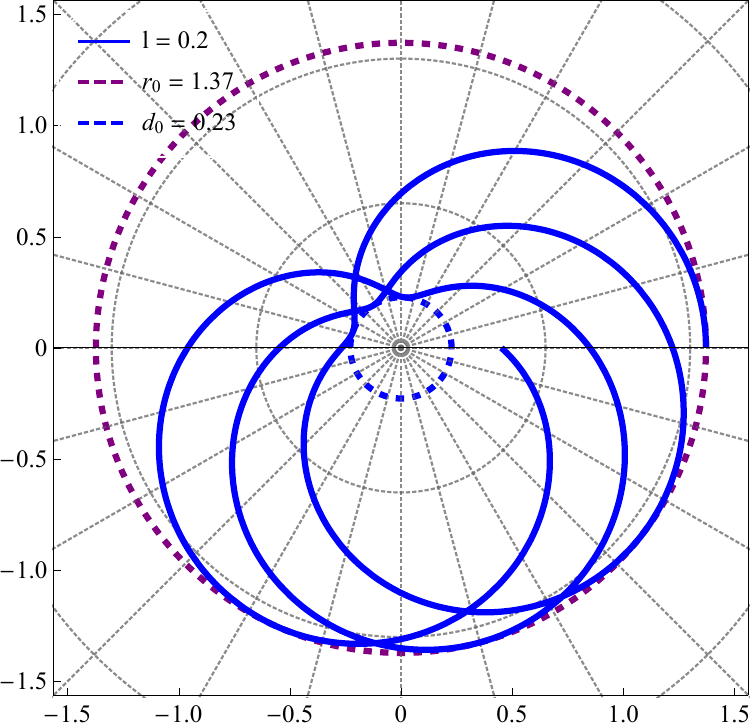}
	\end{center}
	\caption{The Lima\c{c}on of Pascal with precession   (light blue line), with $E_{\ell}= 0.645$. The dashed blue curve indicates the event horizon, while the dashed purple curve indicates $d_0$. We have fixed  $M = 1$,  $Q=0.5$ and $L = 10$.}
	\label{limazon3}
\end{figure}

It is worth mentioning that the analog geodesic in four-dimensional RN--AdS corresponds to the lima\c{c}on of Pascal \cite{Villanueva:2013zta}. Also, when the spacetime is the five-dimensional Schwarzschild–anti-de Sitter spacetime this geodesic is given by $r = 2 M \cos \phi$, which describes a circumference with radius $M$ that is analogous to the cardioid geodesics found in four-dimensional Schwarzschild–anti-de Sitter spacetime \cite{Cruz:2004ts}, and the Hippopede of Proclus geodesic when the spacetime is five-dimensional RN--AdS \cite{Gonzalez:2020zfd}.

The lima\c{c}on of Pascal with precession belongs to the null geodesic structure, within which light deflection stands as the most significant observational evidence. Other geodesics, such as critical trajectories and second class trajectories, which complete the structure, remain unobservable at present. The lima\c{c}on of Pascal with precession, together with other exotic geodesics mentioned such as the cardioid, Pascal’s limaçon, and Proclus’s hippopede, arise in spherically symmetric anti-de Sitter (AdS) spacetimes. Consequently, their physical relevance is closely tied to the existence of a negative cosmological constant. The existence of a negative cosmological constant (i.e., an anti–de Sitter-type vacuum in the dark energy sector), complemented by a quintessence field have been studied, for instance in Refs. \cite{Sen:2021wld, Menci:2024rbq}, which suggest that a 
$\Lambda<0$, which can naturally arise in string theory frameworks, is not ruled out by current cosmological observations.

%\newpage
\subsubsection{Critical trajectories}

In the case of $b = b_u$, the particles can be confined on unstable circular orbits of radius $r_u$. This kind of motion is indeed ramified into two cases; critical trajectories of the first kind (CFK) in which the particles come from a distant position $r_i$ to $r_u$ ($r_i > r_u$) and those of the second kind (CSK) in which the particles start from an initial point $d_i$ in the vicinity of $r_u$ ($d_i < r_u$) and then tend to this radius by spiraling. We obtain the following equations of motion for the aforementioned trajectories:

\begin{equation}\label{criti1}
r_{\text{CFK}}(\phi)=r_u+{2(r_u-r_3)(r_u-r_4)\over (r_3-r_4)\cosh(\kappa_u\,\phi+\varphi_{\infty})+r_3+r_4-2r_u} \,,
\end{equation}%
where 
\begin{eqnarray}
    \kappa_u&=&{\sqrt{(r_u-r_3)(r_u-r_4)}\over\mathcal{ B}_u}\,, \\
    \varphi_{\infty}&=&\cosh^{-1}\left({2r_u-r_3-r_4\over r_3-r_4 } \right)\,,
\end{eqnarray}
for the first kind, while for the second kind  
 \begin{equation}\label{criti2}
 r_{\text{CSK}}(\phi)=r_u-{2(r_u-r_3)(r_u-r_4)\over (r_3-r_4)\cosh(\kappa_u\,\phi)-r_3-r_4+2r_u} \,.
 \end{equation}%

\begin{figure}[!h]
	\begin{center}
		\includegraphics[width=65mm]{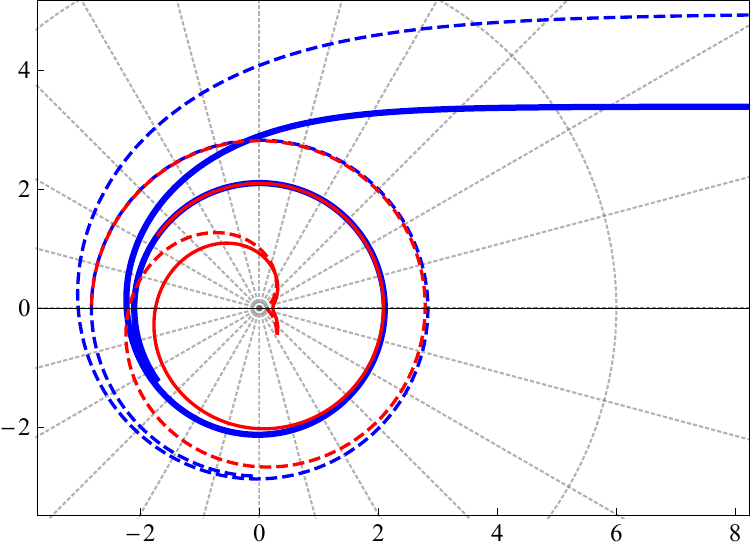}
	\end{center}
	\caption{The critical trajectories  plotted for $M = 1$,  $Q=0.5$, $\Lambda=0$ and $L = 10$. Blue  line for CFK and red  line for CSK trajectories. Dashed curves for $l=0$, $E^2=4.052$ and $r_u =2.823$. Solid curves for $l=0.2$, $E^2=8.757$, and $r_u=2.103$.}
	\label{criticas}
\end{figure}

Fig.~\ref{criticas} displays the critical null trajectories associated with the unstable circular photon orbit. 
The blue curve corresponds to the critical trajectory of the first kind Eq. (\ref{criti1}), where the photon starts from the outer region and asymptotically approaches the circular orbit, while the red curve represents the critical trajectory of the second kind Eq. (\ref{criti2}), where the photon starts from the inner region and also tends asymptotically to the same radius. 
The dashed curves show the RN--(A)dS reference case, obtained for $l=0$, whereas the continuous curves correspond to the Lorentz--violating KR geometry.

The comparison with the $l=0$ case shows that the KR parameter produces a clear inward displacement of the critical structure. 
For the parameters used in the figure, the RN--(A)dS reference orbit has a larger photon-sphere radius, while the KR-deformed case has a smaller value of $r_u$. 
Consequently, both the CFK and CSK trajectories spiral around a more compact circular orbit when the Lorentz--violating parameter is turned on. 
This behavior is consistent with the analysis of the effective potential: increasing $l$ shifts the maximum of the potential toward smaller radial distances and modifies the critical energy and impact parameter associated with the unstable circular null orbit.

It is important to stress that these trajectories are not ordinary deflection or capture orbits. 
They represent the separatrix between photons that escape after being deflected and photons that plunge into the black hole. 
Therefore, the shift observed in Fig.~\ref{criticas} means that the KR background changes the boundary between the deflection and capture regions. 
In this sense, the Lorentz-violating parameter affects not only the location of the photon sphere, but also the critical geodesic structure that separates qualitatively different classes of null motion.

Unlike the first-kind and second-kind non-critical trajectories, the critical curves are controlled mainly by the local properties of the effective potential near its maximum. 
Since the radial position of this maximum is independent of the cosmological constant, the critical trajectories have the same radial location for the asymptotically flat, dS and AdS sectors once $M$, $Q$ and $l$ are fixed. 
The cosmological constant changes the corresponding energy threshold, but it does not shift the critical radius. 
Thus, Fig.~\ref{criticas} isolates the effect of the KR deformation on the critical photon orbits, using the RN--(A)dS case as the natural benchmark.

\bigskip

\subsubsection{Capture zone}
\begin{figure}[H]
	\begin{center}
		\includegraphics[width=65mm]{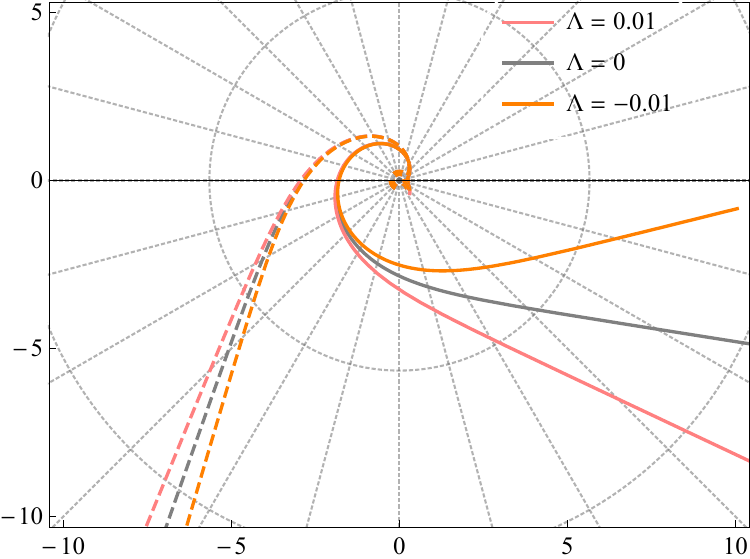}
	\end{center}
    \caption{
Representative trajectories in the capture zone. Incoming photons with $b<b_u$ plunge into the horizon, while the time-reversed branch can be interpreted as an escaping trajectory. We have fixed $M=1$, $Q=0.5$, $L=10$ and $E^2=10$. Dashed curves correspond to $l=0$, and solid curves to $l=0.2$.
}
	\label{captura}
\end{figure}

The photons with an impact parameter smaller than the critical one ($b<b_u$), which are in the capture zone, can plunge into the horizon or escape to infinity, with a cross section given by Eq. (\ref{sigma}). So, by manipulating Eq. (\ref{n11}), and considering $b<b_u$, we show the trajectories of the particles in the capture zone in Fig. \ref{captura}. This figure displays representative trajectories in the capture zone, corresponding to photons with impact parameter smaller than the critical one, $b<b_u$. 
The continuous curves correspond to the Lorentz-violating KR geometry with $l=0.2$, while the dashed curves represent the RN--(A)dS reference case, obtained in the limit $l=0$. 
Therefore, the separation between the continuous and dashed curves measures the deformation of the capture trajectories induced by the KR background.

For incoming photons from the external region, the condition $b<b_u$ implies that there is no external turning point capable of sending the photon back to infinity. 
Thus, the photon is captured by the black hole and eventually crosses the event horizon. 
The time-reversed branch may be interpreted as an escaping trajectory starting from the near-horizon region, but the physically relevant incoming branch belongs to the plunging sector. 
In this sense, these curves are qualitatively different from the first-kind deflection trajectories, where photons reach a turning point outside the photon sphere and escape.

The comparison with the dashed RN--(A)dS trajectories shows that the KR parameter changes the angular structure of the capture process. 
For the same values of $M$, $Q$, $L$ and $E$, the KR deformation reduces the critical impact parameter with respect to the $l=0$ case. 
Consequently, the fixed impact parameter used in the figure lies closer to the critical value in the KR geometry. 
This explains why the continuous curves can display a stronger angular bending before crossing the horizon: the photon approaches more closely the critical separatrix between deflection and capture. 
Thus, even within the capture zone, the Lorentz--violating parameter modifies how long the photon remains in the strong-field region before plunging.

The cosmological constant further changes the global shape of the capture trajectories. 
For fixed $l$, the de Sitter, asymptotically flat and anti-de Sitter branches correspond to different critical impact parameters and different angular openings of the plunging orbit. 
In particular, the AdS branch is closer to the critical regime for the parameters used here, producing a more pronounced bending before capture, while the dS branch gives a comparatively more direct plunging trajectory. 
Therefore, Fig.~\ref{captura} shows that the capture zone is affected by two complementary mechanisms: the KR parameter controls the deviation from the RN--(A)dS reference geometry, whereas the cosmological constant controls the global opening of the plunging trajectories.

\bigskip

\section{Observational Effects}
\label{OBE}

In this section we analyze observational effects associated with photon propagation in the charged KR black-hole spacetime, emphasizing the deviations induced by the Lorentz--violating parameter with respect to the RN--(A)dS limit.

\subsection{Deflection of light}

The process of light deflection, or the so-called gravitational lensing, can theoretically be approached by means of the geodesic equations for the light rays (null geodesics). 
Accordingly, the first-order, angular equation of motion for the light rays (i.e. photons as test particles) passing near the black hole is given by \eqref{em1}.
Performing the change of variable $r=1/u$, the above equation yields

\begin{equation}\label{ue}
	\left(\frac{ du}{d\phi}\right)^2=\frac{1}{b^2}+ \frac{\Lambda }{3(1-l)} - \frac{u^{2}}{(1-l)}
	+2Mu^3-\frac{Q^2\,u^{4}}{(1-l)^2}\,,
\end{equation}
that reduces to the standard Schwarzschild equation of light deflection in the limit of  $l \rightarrow 0$, $\Lambda \rightarrow 0$ and $Q	\rightarrow  0$.
Differentiating Eq. \eqref{ue} with respect to $\phi$, gives

\begin{equation}
	u^{\prime\prime}+u =-\frac{l\,u}{(1-l)}
	+3Mu^2-\frac{2Q^2\,u^{3}}{(1-l)^2}\,.
\end{equation}
Following the procedure established in Ref. \cite{Straumann}, we obtain
\begin{eqnarray}
u &=&\frac{1}{b}\sin\phi-\frac{\sqrt{2}\,l}{2\,(1-l)\,b}+{ 3M\over 2\,b^2}-\frac{\sqrt{2}\,Q^2}{2\,(1-l)^2\,b^3}
\\
\notag && +\left(-\frac{\sqrt{2}\,l}{12\,(1-l)\,b}+{ M\over 2\,b^2}-\frac{\sqrt{2}\,Q^2}{4\,(1-l)^2\,b^3}\right)\cos(2\phi)\,.
\end{eqnarray}
Note that, $u \rightarrow 0$ results in $\phi\rightarrow\phi_{\infty}$, with
\begin{equation}
	-\phi_{\infty}=-\frac{7\sqrt{2}\,l}{12\,(1-l)}+{ 2M\over \,b}-\frac{3\sqrt{2}\,Q^2}{4\,(1-l)^2\,b^2}\,.
\end{equation}
The deflection angle of the light rays passing the black hole is, therefore, obtained as
\begin{equation}\label{GB1}
	\hat{\alpha} = 2\left |-\phi_{\infty}\right | = -\frac{7\sqrt{2}\,l}{6\,(1-l)}+{ 4M\over \,b}-\frac{3\sqrt{2}\,Q^2}{2\,(1-l)^2\,b^2}\,.
\end{equation}
In the limit $l\to0$, Eq.~(\ref{GB1}) reduces to the
RN result, while the Schwarzschild expression
$\hat{\alpha}_{\rm Sch}=4M/b$ is recovered further by taking
$Q\to0$. In this weak-field expansion and with the adopted definition of the bending angle, the cosmological constant does not contribute explicitly to Eq.~(\ref{GB1}), as was pointed out in Ref. \cite{Kagramanova:2006ax}. However, as discussed in Sec.~\ref{NGS}, $\Lambda$ affects the global null-geodesic structure through the anomalous impact parameter, the turning points and the allowed radial domain. 
Assuming the Sun as the central massive object, the deflection angle predicted by the Schwarzschild solution is given by
$\hat{\alpha}_{\mathrm{Sch}} = \frac{4M_{\odot}}{R_{\odot}}$
which, in arcsecond, reads
$
\hat{\delta}_{\mathrm{Sch}} = \frac{3600 \cdot 180}{\pi} \hat{\alpha}_{\mathrm{Sch}}=1.75092 \, \text{arcsec}\,. 
$
Observations near the Sun report measured values of
$\hat{\alpha}_P = 1.7520\, \text{arcsec}$ (prograde) and $\hat{\alpha}_R = 1.7519\, \text{arcsec}$ (retrograde)~\cite{Roy:2019ijp,Fathi:2025byw}. Therefore, by identifying the theoretical prediction with the observational value, we obtain the following constraints:
\begin{align}
\hat{\delta}_{\mathrm{SH}} = \hat{\alpha}_P &\quad \Rightarrow \quad \abs l \leq 3.2\times 10^{-9}\,, \\
\hat{\delta}_{\mathrm{SH}} = \hat{\alpha}_R &\quad \Rightarrow \quad \abs l \leq 2.9\times 10^{-9}\,.
\end{align}
where $\hat{\delta}_{\mathrm{SH}} = \frac{3600 \cdot 180}{\pi} \hat{\alpha}$.

\subsection{Gravitational redshift}

Since the KR black-hole spacetime is stationary, there exists a timelike Killing vector. Therefore, in coordinates adapted to this symmetry, the ratio between the measured frequencies of a photon emitted at $r_0$ and received at $r$ is given by \cite{Kagramanova:2006ax} 
\begin{equation}
{\nu \over \nu_0}=\sqrt{\frac{g_{00}(r)}{g_{00}(r_0)}}\,,
\end{equation}
for  $M/r\ll1$, $Q^2/r^2\ll1$ and $|\Lambda|r^2\ll1$, 
%$M/r \ll 1$ and $Q/r\ll1$, 
the above expression yields 
\begin{eqnarray}
{\nu \over \nu_0}\approx&1&+ M (1-l)\left( \frac{1}{r_0}-\frac{1}{r}\right)-\frac{Q^2}{2(1-l)}\left(\frac{1}{r_0^2}-\frac{1}{r^2}\right)+ \,\nonumber\\
&+&\frac{\Lambda}{6}\left(r_0^2-r^2\right)\,,
\end{eqnarray}
where we have considered the metric function of Eq.~(\ref{FrKRLambda})
\begin{equation}
f(r)=\frac{1}{1-l}
\left[
1-\frac{2M(1-l)}{r}
+\frac{Q^2}{(1-l)r^2}
-\frac{\Lambda r^2}{3}
\right]\,.
\end{equation}
Now, if we consider the limit ${M} 	\rightarrow M_{\oplus}=0.0044346$ m,  and $l	\rightarrow  0$, we recover the classical result for the Schwarzschild dS and AdS spacetime.
The clock can be compared with an accuracy of $10^{-15}$, the H-maser in the GP-A redshift experiment \cite{Vessot:1980zz} reached an accuracy of $10^{-14}$. Therefore, considering that all observations are well described within Einstein's theory with $Q=0$, and $\Lambda=0$, we obtain
\begin{eqnarray}
{\nu \over \nu_0}\approx&1&+ M_{\oplus}\left( \frac{1}{r_0}-\frac{1}{r}\right)-M_{\oplus}l\left( \frac{1}{r_0}-\frac{1}{r}\right)\,.
\end{eqnarray}
So, we conclude that the correction terms to the mass must be $<10^{-14}$. Thus,
\begin{equation}
\abs l \leq 6.154\cdot 10^{-11}\,.
\end{equation}
This estimate assumes that the mass parameter is fixed independently from the redshift measurement. In the Solar-System limit $Q=0$ and $\Lambda=0$, the leading KR correction enters through the combination $M(1-l)$, so that gravitational redshift alone cannot fully disentangle $l$ from a redefinition of the effective gravitational mass. Here, we assume a clock comparison between Earth and a satellite at 15,000 km height, as in Ref. \cite{Kagramanova:2006ax}, where the authors have constrained the cosmological constant.

\subsection{Shapiro time delay}

An interesting relativistic effect in the propagation of light rays is the apparent delay in the time of propagation for a light signal passing near the Sun, which is a relevant correction for astronomical observations and is called the Shapiro time delay. The time delay of Radar Echoes corresponds to the determination of the time delay of radar signals which are transmitted from the Earth through a region near the Sun to another planet or spacecraft and then reflected back to the Earth. The time interval between emission and return of a pulse as measured by a clock on Earth is \cite{Straumann}

\begin{equation}
t_{12}=2\, t(r_1,r_d)+2\, t(r_2,r_d)\,,
\end{equation}
where $r_d$ denotes the distance of closest approach to the Sun. To compute the time delay, we use Eq.~(\ref{g10}) and impose the turning-point condition at $r=r_d$,
\[
\frac{E^2}{L^2}=\frac{f(r_d)}{r_d^2}.
\]
%where $r_d$ as the closest approach to the Sun; now, to calculate the time delay we use Eq. (\ref{g10}), 
%and by considering that
%$dr/dt$ vanishes, thereby
%$\frac{E^2}{L^2}=\frac{f(r_d)}{r_d^2}$.
Thus, the coordinate time that the light needs to go from $r_d$ to $r$ is given by
\begin{equation}
t(r,r_d)=\int_{r_d}^r \frac{dr'}{f(r')\sqrt{1-\frac{r_d^2}{f(r_d)}\frac{f(r')}{r'^2}}}\,.
\end{equation}

So, at first order correction we obtain
\begin{equation}
	t(r, r_d)=\sqrt{r^2-r_d^2}+t_l(r)+t_M(r)+t_Q(r)+t_{\Lambda}(r)\,,
\end{equation}
where
\begin{eqnarray}
t_l(r)&=&-{l\over 1-l}\sqrt{r^2-r_d^2}\,,\\
	t_M(r)&=&M\left(\sqrt{r-r_d\over r+r_d}+2\,\ln\left| \frac{r+ \sqrt{r^2-r_d^2}}{r_d}\right|  \right) \,,\\
	t_{Q}(r)&=&-\frac{3\,Q^2}{2\,(1-l)^2\,r_d}\sec^{-1}\left({r\over r_d}\right)\,,\\
	t_{\Lambda}(r)&=&{\Lambda\over 18\,(1-l)}\left(2\,r^2+r^2_d\right)\sqrt{r^2-r_d^2}\,.
\end{eqnarray}
Therefore, for the circuit from point 1 to point 2 and back, the delay in the coordinate time is
\begin{equation}
	\Delta t = 2\left[t(r_1, r_d)+t(r_2,r_d)-\sqrt{r_1^2-r_d^2}-\sqrt{r_2^2-r_d^2}\right]\,,
\end{equation}
namely,
\begin{eqnarray}
	\notag
	\Delta t &=&2\left[t_{l}(r_1)+t_{l}(r_2)+t_M(r_1)+t_M(r_2)\right]+\\
	&& +2\left[t_{Q}(r_1)+t_{Q}(r_2)+t_{\Lambda}(r_1)+t_{\Lambda}(r_2)\right]\,.
\end{eqnarray}
Now, for a round trip in the solar system, we have ($r_d <<r_1,r_2$)
\begin{eqnarray}
	\label{deltat}
	\notag
	\Delta t &\approx& -{2\,l \over 1-l}(r_1+r_2)+4M\left(1+ \ln\left| \frac{4r_1r_2}{r_d^2}\right| \right) + \\\notag
	&& -\frac{3\,{Q}^2}{(1-l)^2\,r_d}\Bigg[\sec^{-1}\left({r_1\over r_d}\right)+\sec^{-1}\left({r_2\over r_d}\right)\Bigg]+\\
	&&+\frac{2\Lambda}{9\,(1-l)}\left(r_1^3+r_2^3\right)\,.
\end{eqnarray}

Note that if we consider the limit ${M} 	\rightarrow M_{\odot}$ and $\Lambda 	\rightarrow  0$, we recover the classical result of GR; that is, $\Delta t_{\text{GR}}=4M_{\odot}\left[ 1+ \ln\left(\frac{4r_1r_2}{r_d^2}\right)\right]$. For a round trip from Earth to Mars and back, we get (for $r_d \ll r_1 , r_2$), where $r_1 \approx r_2=2.25\times 10^{11}$ m is the average  Earth-Mars distance. Taking $r_d$ as the distance of closest approach to the Sun, approximately the solar radius ($R_{\odot} \approx %696000000
 6.960\times 10^{8}$ m) plus the solar corona ($  \sim 10^{9}$ m),  $r_d \approx1.696\times 10^{9}$ m, the time delay is 
 ${\Delta t_{\text{GR}} \over c}\approx 240\; \mu\,s\,.$
 To give an idea of the experimental possibilities, we mention that the error in the time measurement of a circuit during the Viking mission was only about $10\, ns$ \cite{Straumann}.  If the Lorentz--violating term contributes, then
  \begin{eqnarray}
     \Delta t_{l}=-{2\,l \over 1-l}(r_1+r_2)\,.
 \end{eqnarray}
 For a round trip from Earth to Mars and back, we get (for $r_d \ll r_1 , r_2$)
 \begin{eqnarray}
 {\Delta t_{l} \over c} =-{4\,l\,r_1 \over c\,(1-l)}\,.
 \end{eqnarray}
Requiring this correction to be smaller than the Viking timing accuracy,
\[
\left|\frac{\Delta t_l}{c}\right| \leq 10\,{\rm ns}=10^{-8}\,{\rm s}\,,
\]
yields the illustrative estimate
\[
\abs l \leq 3.3\times10^{-12}\,.
\]

%it does so that $|{\Delta t_{l} \over c} |\leq 10\, ns=10^{-8}\,s$, or	$l \leq 3.3\times 10^ {-12}$.

%\newpage
\section{Lyapunov exponents}
\label{LE}

Lyapunov exponents provide a quantitative measure of the instability of circular geodesics. In the present context, they characterize the rate at which two nearby null trajectories separate from each other in the vicinity of the unstable circular photon orbit. Therefore, they are directly related to the instability timescale of the photon sphere and, in the eikonal limit, to the imaginary part of the quasinormal-mode spectrum \cite{Cardoso:2008bp}.

For the static and spherically symmetric line element (\ref{metricKR})
the radius of the unstable circular null orbit is given by (\ref{rul}).
The existence of this orbit requires
\begin{equation}
9M^2(1-l)^3-8Q^2\geq 0\,.
\end{equation}
As already observed from the effective-potential analysis, $r_u$ is independent of the cosmological constant. This happens because the cosmological constant term contributes to $V_{\rm eff}$ as a constant shift for the null motion and, therefore, does not modify the extremum condition. However, $\Lambda$ affects the value of the effective potential at maximum and, consequently, the critical energy and the impact parameter.

Following the standard phase-space analysis \cite{Cardoso:2008bp,Pradhan:2012rkk,Pradhan:2013bli}, the Lyapunov exponent associated with the circular null orbit can be written as
\begin{equation}
\lambda
=
\frac{1}{\sqrt{2}}
\sqrt{
-\frac{r_u^2 f(r_u)}{L^2}
V_{\rm eff}''(r_u)
}\,.
\end{equation}
Equivalently, using $V_{\rm eff}=L^2 f/r^2$, one obtains the purely geometrical expression
\begin{equation}
\lambda^2
=
-\frac{f(r_u)}{2r_u^2}
\left[
r_u^2 f''(r_u)-4r_u f'(r_u)+6f(r_u)
\right]\,.
\end{equation}
For the KR black hole metric given above, this yields
\begin{equation}
\lambda^2
=
-\frac{f(r_u)}{2}
\left[
\frac{6}{(1-l)r_u^2}
-\frac{24M}{r_u^3}
+
\frac{20Q^2}{(1-l)^2r_u^4}
\right]\,.
\end{equation}
Using the circular orbit condition, the previous expression can be cast in the compact form
\begin{equation}
\lambda^2
=
\frac{
\left[2r_u-3M(1-l)\right]
\left[3r_u-3M(1-l)-2\Lambda r_u^3\right]
}
{6(1-l)^2 r_u^4}\,.
\end{equation}
This expression explicitly shows that, although the location of the circular photon orbit does not depend on $\Lambda$, the instability timescale does. Therefore, the cosmological constant changes the rate at which nearby null trajectories separate from the unstable circular orbit.

In the asymptotically flat limit $\Lambda=0$, the Lyapunov exponent reduces to
\begin{equation}
\lambda^2_{\Lambda=0}
=
\frac{
\left[2r_u-3M(1-l)\right]
\left[r_u-M(1-l)\right]
}
{2(1-l)^2 r_u^4}\,.
\end{equation}
In turn, in the limit $l\rightarrow 0$, one recovers the corresponding RN--(A)dS expression \cite{Pradhan:2012rkk,Gim:2019rkl,Guo:2022kio},
\begin{equation}
r_u=
\frac{3M}{2}
+
\sqrt{\frac{9M^2}{4}-2Q^2}\,,
\end{equation}
and
\begin{equation}
\lambda^2
=
\frac{
(2r_u-3M)(3r_u-3M-2\Lambda r_u^3)
}
{6r_u^4}\,.
\end{equation}
For $Q=0$ and $l=0$, we further obtain $r_u=3M$ and
\begin{equation}
\lambda^2
=
\frac{1}{27M^2}
-\frac{\Lambda}{3},
\end{equation}
which is the standard Schwarzschild--(A)dS result for the instability of the photon sphere \cite{Cardoso:2008bp}.

The sign of $\lambda^2$ determines the character of the circular null orbit. When $\lambda^2>0$, the orbit is unstable; when $\lambda^2=0$, the orbit is marginal; and when $\lambda^2<0$, the orbit is stable under radial perturbations. For the outer photon orbit considered here, the factor $2r_u-3M(1-l)$ is non-negative whenever the circular orbit exists. Hence, the instability is controlled by the second factor,
\begin{equation}
3r_u-3M(1-l)-2\Lambda r_u^3\,.
\end{equation}
For $\Lambda\leq 0$, this factor is enhanced, and the circular null orbit remains unstable. For positive $\Lambda$, the cosmological constant decreases the value of $\lambda^2$ and  increases 
the instability timescale, consistently with the presence of a cosmological horizon and the modification of the allowed photon region.

\begin{figure}[!h]%
	\begin{center}
	\includegraphics[width=65mm]{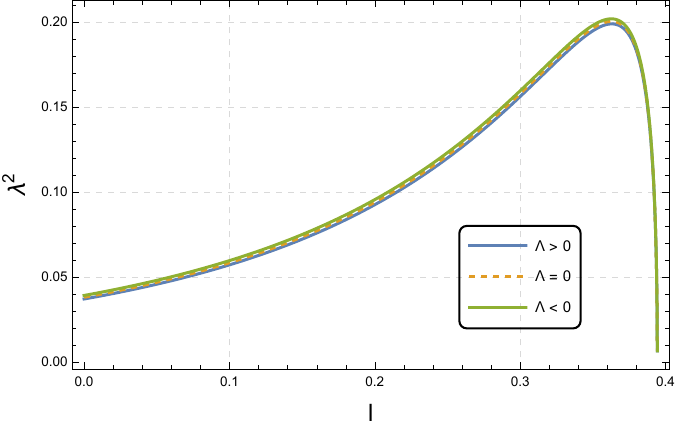}
    \includegraphics[width=65mm]{figLl.pdf}
	\end{center}%
\caption{
Lyapunov exponent squared $\lambda^2$ associated with the unstable circular null orbit of the charged KR black hole. 
The upper panel shows $\lambda^2$ as a function of the cosmological length scale $\ell$, defined by $\ell^2=|3/\Lambda|$, for a fixed value of the Lorentz--violating parameter $l=0.2$. 
The lower panel shows $\lambda^2$ as a function of the Lorentz--violating parameter $l$, for asymptotically de Sitter $(\Lambda>0)$, asymptotically flat $(\Lambda=0)$ and asymptotically anti-de Sitter $(\Lambda<0)$ spacetimes with $\ell=30$. 
In both panels we have fixed $M=1$ and $Q=0.5$.}
	\label{Lyapunov}
\end{figure}

In Fig.~\ref{Lyapunov} we show the behavior of the Lyapunov exponent squared associated with the unstable circular null orbit. The upper panel displays $\lambda^2$ as a function of the cosmological length scale $\ell$, with $\ell^2=|3/\Lambda|$. For small values of $\ell$, the de Sitter and anti-de Sitter branches depart significantly from the asymptotically flat result. In particular, the AdS branch gives larger values of $\lambda^2$, indicating a stronger instability of the circular photon orbit, whereas the dS branch gives smaller values, showing that a positive cosmological constant tends to reduce the instability timescale. As $\ell$ increases, both branches approach the flat case, consistent with the limit $\Lambda\rightarrow 0$.

The lower panel of Fig.~\ref{Lyapunov} isolates the effect of the Lorentz--violating parameter $l$. We observe that $\lambda^2$ increases as $l$ grows from zero, meaning that the KR background  initially enhances the instability of the photon sphere. This enhancement is common to the three asymptotic sectors, although the AdS case remains slightly above the flat and dS cases, in agreement with the behavior observed in the upper panel. However, the growth of $\lambda^2$ is not monotonic throughout the interval: after reaching a maximum, $\lambda^2$ decreases rapidly as the upper limit of the allowed parameter region approaches. This behavior is a consequence of the dependence of the photon-sphere radius $r_u$ on $l$ and of the condition required for the existence of real circular null orbits. Therefore, the Lyapunov exponent provides a sensitive diagnostic of how spontaneous Lorentz symmetry breaking modifies the stability properties of photon motion in charged KR black hole spacetimes.

It is also useful to compare the Lyapunov exponent with the surface gravity at the black hole horizon. For a horizon $r_h$ that satisfies $f(r_h)=0$, the surface gravity is
\begin{equation}
\kappa_h
=
\frac{1}{2}\left|f'(r_h)\right|
=
\left|
\frac{M}{r_h^2}
-\frac{Q^2}{(1-l)^2r_h^3}
-\frac{\Lambda r_h}{3(1-l)}
\right|\,.
\end{equation}
The chaos bound proposed by Maldacena, Shenker and Stanford reads \cite{Maldacena:2015waa}
\begin{equation}
\lambda \leq 2\pi T\,,
\end{equation}
and, using the black-hole relation $T=\kappa_h/(2\pi)$, can be written as
\begin{equation}
\lambda \leq \kappa_h\,.
\end{equation}
In the present geometry, both $\lambda$ and $\kappa_h$ are modified by the Lorentz--violating KR parameter. Therefore, the comparison between these two quantities provides a useful diagnostic of how the KR background affects the instability of photon motion and its relation with the near-horizon thermodynamic scale.

To make this comparison quantitative, we define the dimensionless ratio
\begin{equation}
\chi\equiv\frac{\lambda}{\kappa_h}.
\end{equation}
The condition $\lambda\leq\kappa_h$ is then equivalent to $\chi\leq1$.
We evaluate this ratio numerically for the same set of parameters used in Fig.~\ref{Lyapunov}, namely $M=1$, $Q=0.5$ and $\ell=30$, with $\Lambda=\pm3/\ell^2$ for the dS and AdS branches. For moderate values of the Lorentz-violating parameter, the ratio remains below unity. For instance, at $l=0.2$ one obtains
\begin{equation}
\chi_{\rm dS}\simeq0.808\,,\qquad
\chi_{0}\simeq0.807\,,\qquad
\chi_{\rm AdS}\simeq0.806\,.
\end{equation}
Thus, in this regime the geodesic instability scale is smaller than the near-horizon scale set by the surface gravity.

However, as $l$ approaches the upper region allowed by the existence of the black-hole horizon, the surface gravity decreases rapidly and the ratio $\chi$ grows. For the same parameters, the crossing $\chi=1$ occurs at
\begin{equation}
l_{\rm c}^{\rm dS}\simeq0.343997\,,\qquad
l_{\rm c}^{0}\simeq0.343894\,,\qquad
l_{\rm c}^{\rm AdS}\simeq0.343791\,.
\end{equation}
For larger values of $l$, but still before the limiting horizon configuration is reached, one finds $\chi>1$. This indicates that the instability timescale of the photon sphere may become shorter than the horizon thermal scale in the near-extremal KR regime. Therefore, the comparison between $\lambda$ and $\kappa_h$ should be understood here as a geometrical diagnostic of the relative strength of photon-sphere instability, rather than as a direct test of the quantum chaos bound.

\section{Concluding comments}
\label{conclution}

In this work we have investigated the null geodesic structure of four-dimensional charged KR black holes in the presence of a cosmological constant. The geometry is characterized by a Lorentz--violating parameter $l$ induced by the non-vanishing vacuum expectation value of the KR field. This parameter continuously deforms the RN--(A)dS family, which is recovered in the limit $l\rightarrow0$. Therefore, the results obtained in this work can be interpreted as Lorentz--violating departures from the standard RN, RN--dS and RN--AdS geometries.

We first analyzed radial null trajectories. For photons with vanishing angular momentum, the radial motion in terms of the affine parameter keeps the same simple form in the asymptotically flat, de Sitter and anti-de Sitter sectors. However, the coordinate time is sensitive to the horizon structure of the corresponding geometry. In the RN limit, one recovers the usual behavior in which an external observer measures an infinite coordinate time for photons approaching the event horizon. The KR deformation preserves this qualitative behavior, but shifts the location of the horizons and changes the coordinate-time scale. In the de Sitter branch, the RN--dS causal structure with an event horizon and a cosmological horizon is maintained, while the KR parameter changes the size of the static region. In the AdS branch, where no cosmological horizon is present, the KR deformation modifies the event-horizon position and the approach to the asymptotic AdS region.

For photons with non-vanishing angular momentum, we derived the effective potential and analyzed the corresponding circular and non-circular trajectories. The radius of the unstable circular null orbit is modified by the electric charge and by the Lorentz--violating parameter, but it remains independent of the cosmological constant. This result generalizes the known property of RN--(A)dS spacetimes, where the cosmological constant changes the value of the effective potential at its maximum but does not shift the photon-sphere radius. Thus, in the present geometry, the KR parameter controls the local structure of the photon sphere, while the cosmological constant controls the energetic scale associated with critical motion.

The comparison with the limiting RN, RN--dS and RN--AdS cases is particularly useful for interpreting the photon trajectories. In the asymptotically flat sector, the $l=0$ curve corresponds to the standard RN deflection and capture structure, whereas the curves with $l>0$ show how the KR background modifies the closest approach distance, the critical impact parameter and the capture cross section. In the de Sitter sector, the reference solution is RN--dS, and the KR parameter deforms the photon trajectories inside the finite static patch bounded by the event and cosmological horizons. In the anti-de Sitter sector, the $l=0$ solution corresponds to RN--AdS, where the confining character of the geometry allows second-kind trajectories and special limiting orbits. The KR parameter preserves this qualitative classification but changes the radial domain and angular structure of the trajectories.

We obtained analytic expressions for the angular photon trajectories in terms of Weierstra{\ss} elliptic functions. These solutions allowed us to describe first-kind deflection trajectories, second-kind plunging trajectories, critical trajectories and capture orbits in a unified way. In the first-kind sector, the RN--(A)dS curves provide the natural reference trajectories. Turning on the KR parameter enhances the effective deflection barrier and shifts the turning point of the orbit with respect to the corresponding RN--(A)dS case. For sufficiently large values of $l$, this produces an outward-bending behavior in the deflection trajectories. This effect should not be interpreted as a genuine repulsive gravitational force, but rather as an effective repulsive deflection induced by the Lorentz--violating deformation of the null effective potential.

In the second-kind sector, the KR deformation also produces clear deviations from the RN--(A)dS reference case. These trajectories do not correspond to photons escaping from infinity; rather, they describe photons in the inner allowed region that reach a return point and then plunge into the event horizon. The comparison with the $l=0$ curves shows that the KR parameter modifies the inner potential barrier and changes the radial extension of the plunging trajectories. This demonstrates that Lorentz symmetry breaking affects not only the photon sphere and the usual lensing trajectories, but also the non-critical inner part of the null geodesic structure.

A distinctive feature appears in the AdS branch, where a special limiting trajectory exists when the anomalous impact parameter diverges. In the RN--AdS limit this trajectory reduces to the known lima\c{c}on of Pascal with precession null geodesic. In the KR geometry, the same orbit acquires a deformation controlled by the Lorentz--violating parameter. Increasing $l$ changes both the radial size of the orbit and its angular behavior, producing a precessing lima\c{c}on-type trajectory. This provides a clean analytic signal of how the KR background deforms the RN--AdS geodesic structure.

The bending of light was analyzed by combining the exact geodesic solution with the qualitative structure of the effective potential. The de Sitter, asymptotically flat, and anti-de Sitter branches exhibit different turning points and different approaches to the critical regime. In the limit $l=0$, these results reproduce the corresponding RN, RN--dS and RN--AdS behavior. For $l>0$, the KR parameter changes the closest approach distance, the critical impact parameter, and the energy threshold for strong deflection. Therefore, the difference between the KR trajectories and the RN--(A)dS reference curves provides a direct measure of the Lorentz--violating deformation.

We have also analyzed weak-field observables associated with photon propagation, namely the bending of light, gravitational redshift and the Shapiro time delay. These effects provide complementary Solar-System probes of deviations from General Relativity. In the present geometry, the weak-field expansion reduces to the standard RN--(A)dS result when $l=0$, while the KR parameter introduces additional corrections in the deflection angle, the frequency shift and the radar time delay. By comparing the weak-deflection expression with Solar-System measurements, we obtained the illustrative bounds $|l|\leq 3.2\times10^{-9}$ and $|l|\leq 2.9\times10^{-9}$ from the prograde and retrograde deflection angles, respectively. The gravitational-redshift estimate gives $|l|\leq 6.154\times10^{-11}$, under the assumption that the mass parameter is fixed independently, while the Shapiro time-delay analysis yields the stronger estimate $|l|\leq 3.3\times10^{-12}$. These constraints should be regarded as weak-field Solar-System estimates rather than model-independent bounds, since they rely on the approximations and observational inputs adopted in each test. Nevertheless, they show that weak-field observables can provide useful complementary information to the strong-field photon-sphere, capture and Lyapunov analyses.

Finally, we studied the instability of the unstable circular null orbit by computing the corresponding Lyapunov exponent. In the RN--(A)dS limit, the result reduces to the standard expression for the instability of charged black hole photon spheres with cosmological constant. The KR parameter modifies this instability scale without destroying the limiting structure. In particular, the cosmological constant does not shift the location of the photon sphere, but it does change the instability timescale: the AdS branch enhances the instability, while the dS branch reduces it. The Lorentz--violating parameter produces a non-trivial modification of the Lyapunov exponent, initially increasing the instability of the photon sphere before the allowed parameter range is approached. Thus, the Lyapunov exponent provides a sensitive diagnostic of the deviation from RN--(A)dS induced by the KR background.

Overall, our results show that charged KR black holes provide a controlled Lorentz--violating deformation of the RN, RN--dS and RN--AdS geodesic structures. The limiting cases are recovered smoothly when $l\rightarrow0$, while finite values of $l$ leave imprints on the horizon structure, photon sphere, effective potential, critical impact parameter, capture region, deflection trajectories, and instability timescale. The cosmological constant further enriches the structure by separating the asymptotically flat, de Sitter, and anti-de Sitter sectors. These results suggest that null geodesics provide a useful probe of Lorentz--violating gravitational backgrounds and may contribute to the phenomenological study of KR black holes through lensing, shadows, time-delay effects, and quasinormal-mode physics.

A natural continuation of the present work is the study of timelike geodesics in the charged KR black-hole spacetime. 
Such an analysis would complete the geodesic structure of this geometry by complementing the null sector studied here with the motion of massive test particles. 
In particular, it would allow one to investigate bound orbits, stable and unstable circular trajectories, perihelion precession, plunging motion and the corresponding limiting cases associated with the RN, RN--dS and RN--AdS geometries. 
This extension would provide a more complete characterization of how the Lorentz--violating parameter modifies particle dynamics in both the strong-field and weak-field regimes.

\acknowledgments

Y. V. acknowledges support by the Direcci\'on de Investigaci\'on y Desarrollo de la Universidad de La Serena, Grant No. PR25538511.

\end{document}